\documentclass[twocolumn]{autart}
\makeatletter
\def\ps@copyright{\let\@mkboth\@gobbletwo\def\@oddhead{}\let\@evenhead\@oddhead\def\@oddfoot{}\let\@evenfoot\@oddfoot}
\makeatother

\usepackage{graphicx}
\usepackage{amsmath,amsfonts,mathtools}
\usepackage{empheq}
\usepackage{cite}
\usepackage{url}
\usepackage{algorithm}
\usepackage{algpseudocode}
\usepackage{tikz}

\usetikzlibrary{positioning, fit, arrows.meta, calc, 3d, svg.path, fadings}

\let\b\mathbf
\let\bs\boldsymbol
\newcommand{\overbar}[1]{\mkern 1.5mu\overline{\mkern-1.5mu#1\mkern-1.5mu}\mkern 1.5mu}

\newtheorem{assump}{Assumption}
\newtheorem{theorem}{Theorem}
\newtheorem{definition}{Definition}
\newtheorem{problem}{Problem}
\newtheorem{subproblem}{Problem}[problem]

\newtheorem{remark}{Remark}
\newtheorem{lemma}{Lemma}

\begin{document}

\begin{frontmatter}

\title{Safe Autonomous Landing of Rocket-Like Vehicles: A CBF-governed Path Following Approach \thanksref{footnoteinfo}}

\thanks[footnoteinfo]{Corresponding author: C. Silvestre.}

\author[idmec,isr]{Pedro Santos}\ead{pedrodossantos31@tecnico.ulisboa.pt},
\author[macau]{Joel Reis}\ead{joelreis@um.edu.mo},
\author[idmec,isr]{Paulo Oliveira}\ead{paulo.j.oliveira@tecnico.ulisboa.pt},
\author[isr,macau]{Carlos Silvestre}\ead{csilvestre@um.edu.mo}

\address[idmec]{Institute of Mechanical Engineering, IST, University of Lisbon, 1049-001 Lisbon, Portugal}
\address[isr]{Institute of Systems and Robotics, IST, University of Lisbon, 1049-001 Lisbon, Portugal}
\address[macau]{Department of Robotics, University of Macau, Macau SAR, China}


\begin{abstract}
 This paper addresses the safe autonomous landing of rocket-like vehicles under a glideslope constraint. The landing is formulated as a path following problem, and the progression of the path parameter is used to enforce the constraint. We propose a framework in which only the highest-order time derivative of the path parameter of a pre-stabilized system is regulated through an Exponential Control Barrier Function (ECBF)-based quadratic program. The resulting quadratic program admits a closed-form solution and deviates from the nominal progression only when necessary to satisfy the ECBF condition. At the same time, the stability guarantees of the pre-stabilized system are preserved. For a thrust-vector controlled rocket-like vehicle, we design a saturated nonlinear path following law and an ECBF for the glideslope constraint, and establish conditions under which the vehicle reaches the landing site without leaving the safe set. Compared to optimal guidance, our proposed method trades optimality for a closed-form solution with formal guarantees. Flight experiments on an in-house rocket-like aerial platform, including an aggressive landing scenario, and Monte Carlo simulations on a high-fidelity simulator validate the approach.
\end{abstract}

\end{frontmatter}

\section{Introduction}
Powered-descent landing is an enabling technology for reusable launch vehicles and for missions to the surface of the Moon and Mars. The landings of orbital-class boosters and planetary landers have shown that it can be performed autonomously and with high precision \cite{Blackmore2016,Song2020}. The terminal phase remains a demanding control problem: the translational motion is underactuated, since the main engine provides the only significant force and its direction can only be changed by gimbaling the engine or rotating the vehicle \cite{Szmuk2020,Gomez2026}, and the vehicle must reach the landing site with near-zero velocity while satisfying state constraints, such as a glideslope constraint that keeps it away from the terrain.

Powered-descent guidance is commonly cast as a finite-horizon optimal control problem that minimizes fuel consumption and handles several constraints, such as bounds on the thrust magnitude and pointing direction and a glideslope constraint. Convexification allows such problems to be solved reliably on board \cite{Acikmese2007, Acikmese2011, Malyuta2022}, both with simplified point mass and six-degree-of-freedom models \cite{Szmuk2020}. Model predictive control has been used to close the loop around them \cite{Lee2017}. These approaches, however, require the iterative solution of an optimization problem on board, whose convergence is not always guaranteed \cite{Szmuk2020}, and guidance and control are usually designed separately, which makes formal guarantees for the closed loop difficult to obtain, in particular far from the reference trajectory. Moreover, re-solving the propellant-optimal problem until touchdown can lead to unacceptable maneuvers near the landing site, and switching to a globally convergent tracking law in the final part of the descent has been recommended \cite{Lu2023}.

An alternative is to design the position and attitude control jointly, through a nonlinear feedback law with formal stability guarantees for the complete closed loop \cite{Pucci2015,Santos2026}, which also provides the convergent tracking recommended near touchdown. However, such laws do not enforce state constraints, which then have to be added on top of them, either by modifying their input or their reference. In the first approach, control barrier functions (CBFs) enforce state constraints through a quadratic program (QP) that minimally modifies a nominal input \cite{Ames2017,Ames2019}, using exponential \cite{Nguyen2016} or high-order \cite{Xiao2022} CBFs for constraints of high relative degree. When the CBF acts directly on the physical input, the stability guarantees of the nominal design no longer apply in general, and the QP can introduce undesirable asymptotically stable equilibria \cite{Reis2021}. For launch vehicles, a CBF-based filter of the thrust-vector input was found to compromise the stability of the attitude dynamics, and the safety constraints had to be relaxed to preserve it \cite{Gomez2026}. CBFs designed for reduced-order models can instead be enforced through the reference of a tracking controller \cite{Molnar2022}, under conditions on the tracking performance.

In the second approach, reference governors leave the pre-stabilized system unchanged and enforce constraints by manipulating only the reference applied to it \cite{Garone2017}. The explicit reference governor \cite{Nicotra2018} does so in closed form, scaling the motion of the applied reference by a dynamic safety margin (DSM), and has been applied to multirotor aerial vehicles \cite{Tartaglione2024}. DSMs have recently been shown to be CBFs of the system augmented with the reference \cite{Freire2026}, which led to CBF-based reference governors \cite{Nakano2026}. DSMs, however, are typically obtained from Lyapunov sublevel sets of the pre-stabilized system under the assumption of a constant reference, which can be conservative \cite{Nicotra2018} and requires a Lyapunov function of the complete closed loop \cite{Tartaglione2024}. For an underactuated vehicle under a saturated nonlinear control law, such bounds can be loose and slow down the descent more than necessary.

For a landing, the reference is naturally a path to the landing site. In path following, the timing along the path is not fixed in advance, which can remove performance limitations of trajectory tracking and provides an additional degree of freedom \cite{Aguiar2008}. The progression of the path parameter has been used to handle constraints in model predictive path following \cite{Faulwasser2016}, by governors that slow it down based on predictions of the closed loop or on DSM-type barrier functions \cite{Li2023}, and as an additional input of CBF-based QPs, together with the physical input \cite{Zheng2023,Carvalho2026}. These governors rely on predictions or DSMs and can therefore slow down the progression even far from the constraints, whereas the CBF-based formulations modify the physical input and only consider constraints of relative degree one or two.

Building on these ideas, we trade the optimality of the solution for provable stability and safety. The landing is formulated as a path following problem, with a desired path that ends at the landing site with zero velocity, and the design is carried out in two steps. First, a saturated nonlinear path following law \cite{Reis2023,Santos2026} ensures the almost global convergence of the tracking errors to the origin. Second, a path governor enforces a glideslope constraint by modifying only the highest time derivative of the path parameter required by the control law, through an exponential CBF (ECBF) of the constraint function itself, computed along the closed loop. The lower derivatives are obtained by integration, so that the progression remains as smooth as the path following law requires.

This combination distinguishes our approach from the works above in three respects. First, the physical input is never modified, unlike in \cite{Gomez2026,Zheng2023,Carvalho2026}, so that the stability guarantees of the path following law are preserved. Second, safety does not rely on DSMs, predictions or reduced-order models \cite{Nicotra2018,Freire2026,Nakano2026,Li2023,Molnar2022}, and constraints of arbitrary relative degree are handled. Third, the governor is given by a QP with a single constraint and a closed-form solution, which is inactive when the nominal progression satisfies the ECBF condition. In this way, the nominal path and its time derivatives, which could come from the last trajectory computed by an optimal guidance law \cite{Lu2023}, are recovered whenever the governor is inactive. The price is that the validity of the CBF is not guaranteed by construction, as it is for DSM-based CBFs \cite{Freire2026}, and has to be established for the application. The governor is developed for a general class of pre-stabilized systems and then applied to the rocket-like vehicle. As a first step, only the glideslope constraint is considered, whereas optimal control formulations typically also handle bounds on the thrust magnitude and pointing direction. The main contributions of this paper can be summarized as follows.
\begin{enumerate}\renewcommand{\labelenumi}{(\roman{enumi})}
	\item A path governor for a general class of pre-stabilized systems, which enforces a state constraint of arbitrary relative degree by modifying only the highest time derivative of the path parameter, through an ECBF-QP with a closed-form solution. The pre-stabilizing law is never modified, so that stability guarantees are preserved.
	\item A safe landing controller for thrust-vectoring rocket-like vehicles, which combines a saturated nonlinear path following law with an ECBF of relative degree four for a smooth glideslope constraint. Conditions under which the vehicle never leaves the safe set and reaches the landing site are derived.
\end{enumerate}

	Monte Carlo simulations on a high-fidelity simulator and flight experiments on the \textit{QuadRocket} \cite{Santos2026}, including an aggressive entry into the landing phase that emulates the end of an unpowered descent, are used to validate the proposed approach.

\subsection{Notation}
Bold lowercase and uppercase symbols stand for column vectors and matrices, respectively. The symbol $\mathbf{I}$ denotes the identity matrix, and $\b{0}$ denotes a vector/matrix of zeros, both of appropriate dimensions. The transpose operator is denoted by $(\cdot)^\mathsf{T}$. The $n$-dimensional Euclidean space is represented by ${\mathbb{R}}^n$. Given a vector $\b{x} \in \mathbb{R}^n$, its Euclidean norm is defined as $\|\b{x}\| \coloneq \sqrt{\b{x}^\mathsf{T}\b{x}}$. The 2-sphere is defined as $\mathbb{S}^2\coloneq\left\{\b{x}\in\mathbb{R}^3:\b{x}^\mathsf{T}\b{x}=1\right\}$. The space tangent to $\mathbb{S}^2$ at a point $\b{p}\in\mathbb{S}^2$ is denoted by $T_\b{p}\mathbb{S}^2$. The special orthogonal group of order three is defined as $\mathrm{SO}(3)\coloneq\left\{\b{X}\in \mathbb{R}^{3\times3}: \b{X}\b{X}^\mathsf{T}=\b{X}^\mathsf{T} \b{X} = \b{I},\,\text{det}(\b{X})=1\right\}$. The operator $\b{S}(\cdot):\mathbb{R}^3\to \mathbb{R}^{3\times3}$ yields the skew-symmetric matrix such that $\b{S}(\b{x})\b{y} =\b{x} \times \b{y} $ for any $\b{x},\,\b{y}\in \mathbb{R}^3$, where $\times$ denotes the cross product. A function $f$ is of class $\mathcal{C}^n$ if its derivatives $f^{(1)}, f^{(2)}, \ldots, f^{(n)}$ exist and are continuous. The vectors $\b{e}_1,\,\b{e}_2,\,\b{e}_3 \in \mathbb{S}^2$ are orthonormal and form a basis of $\mathbb{R}^3$, such that ${\b{I}}=\left[{\b{e}}_1\;\b{e}_2\;\b{e}_3\right]$. A smooth saturation function $\bs{\sigma}:\mathbb R^3\times(0,\infty)\to\mathbb R^3$ is defined as
\begin{equation}\label{eq:saturation}
\bs{\sigma}(\b{x},s)\coloneq \frac{s\,\b{x}}{\sqrt{\|\b{x}\|^2+s^2}}\,,
\end{equation}
where $s>0$ defines the saturation level. The state-space representation of a single-input chain of $n$ integrators, useful to describe path parameter and barrier function dynamics, is given by $\dot{\b{x}} = \b{A}_n\b{x} + \b{B}_nu\,,$ where $\b{x}\in\mathbb{R}^n$ is the state vector, $u\in\mathbb{R}$ is the input, and $\b{A}_n\in\mathbb{R}^{n\times n}$ and $\b{B}_n\in\mathbb{R}^{n\times 1}$ are constant matrices defined as
\begin{equation}\label{eq:chain_integrators}
	\b{A}_n\coloneq\begin{bmatrix} \b{0} & \b{I} \\ 0 & \b{0}^\mathsf{T} \end{bmatrix}\,,\hspace{10pt}
	\b{B}_n\coloneq \begin{bmatrix} \b{0}\\ 1 \end{bmatrix}\,.
\end{equation}

\section{Problem Formulation and Preliminaries}

\subsection{Mathematical model}
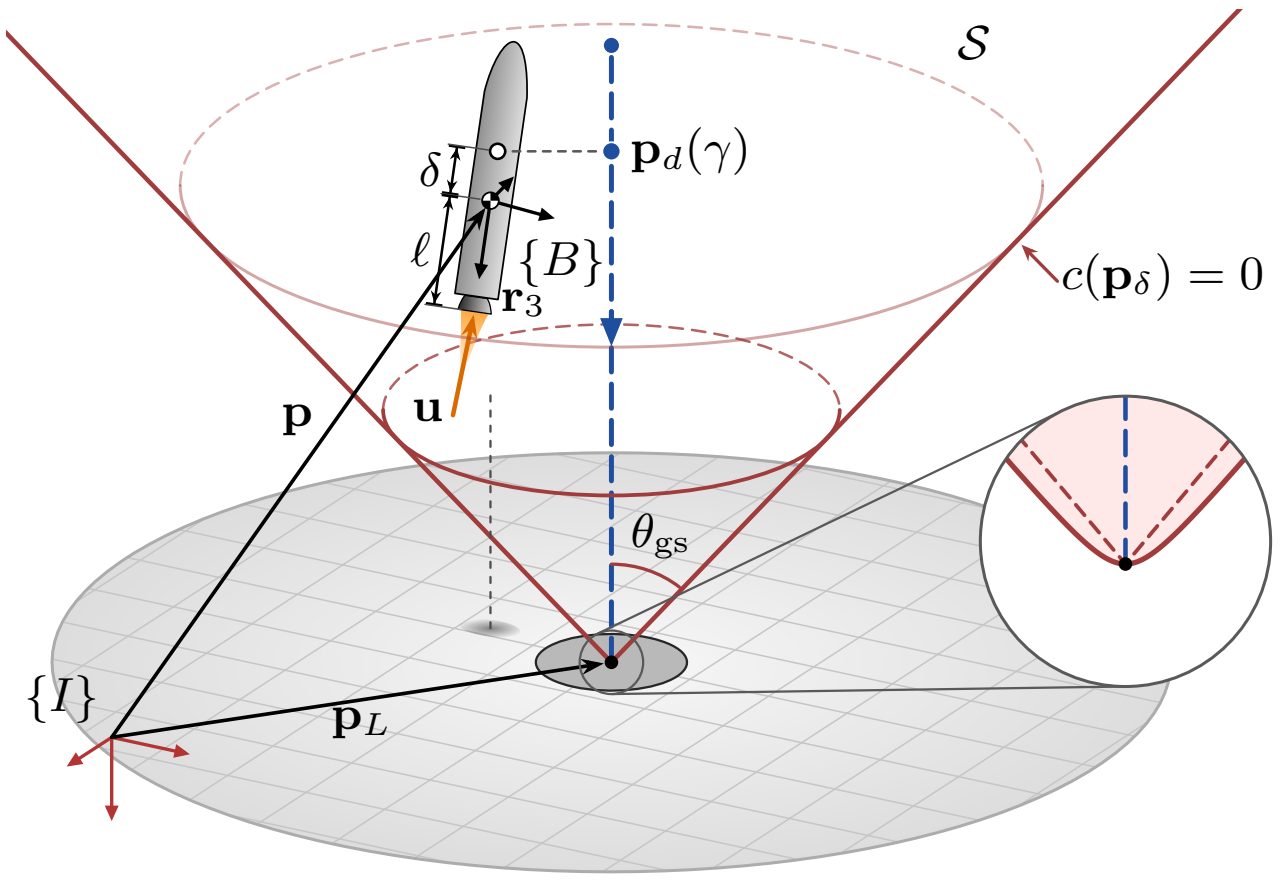
\begin{figure}[!t]
\centering
\tikzfading[name=gsfade,  bottom color=transparent!30, top color=transparent!100]
\tikzfading[name=gsfadeL, bottom color=transparent!0,  top color=transparent!85]
\tikzfading[name=shadowfade, inner color=transparent!0, outer color=transparent!100]
\begin{tikzpicture}[x=1mm, y=1mm, font=\footnotesize, line cap=round, line join=round,
  vec/.style={-{Stealth[length=1.5mm, width=1.05mm]}},
  axis/.style={-{Triangle[length=1.05mm, width=0.85mm]}, line width=0.22mm}]
\pgfmathsetmacro{\el}{22}
\pgfmathsetmacro{\az}{-30}
\pgfmathsetmacro{\thgs}{40}
\pgfmathsetmacro{\hzero}{44}
\pgfmathsetmacro{\Rg}{37}
\pgfmathsetmacro{\rkoff}{8}
\pgfmathsetmacro{\rkalt}{30.5}
\pgfmathsetmacro{\rkdepth}{6}
\pgfmathsetmacro{\rks}{0.46}
\pgfmathsetmacro{\rktau}{8}
\pgfmathsetmacro{\rkkap}{170}
\pgfmathsetmacro{\rkpsi}{90}
\pgfmathsetmacro{\rkdelta}{3.3}
\pgfmathsetmacro{\epsin}{2.4}
\pgfmathsetmacro{\toplev}{34}
\definecolor{gsline}{HTML}{A03C3C}\definecolor{gsfill}{HTML}{FF9999}
\colorlet{gsedge}{gsfill!45!white}\colorlet{gsmid}{gsfill!20!white}
\definecolor{frI}{HTML}{B43333}\definecolor{pathc}{HTML}{1F4E9E}\definecolor{thrc}{HTML}{D86A00}
\definecolor{rkdark}{HTML}{919191}\definecolor{rklight}{HTML}{DEDEDE}\definecolor{flame}{HTML}{FFA93F}
\pgfmathsetmacro{\exx}{cos(\az)}\pgfmathsetmacro{\exy}{sin(\el)*sin(\az)}
\pgfmathsetmacro{\eyx}{sin(\az)}\pgfmathsetmacro{\eyy}{-sin(\el)*cos(\az)}
\pgfmathsetmacro{\ezy}{-cos(\el)}
\coordinate (pL) at (40,16);
\tikzset{view/.style={shift={(pL)}, x={(\exx mm,\exy mm)}, y={(\eyx mm,\eyy mm)}, z={(0mm,\ezy mm)}}}
\useasboundingbox (0,1.6) rectangle (84.35,59.2);
\clip (0,1.6) rectangle (84.35,59.2);
\pgfmathsetmacro{\ts}{asin(tan(\el)*tan(\thgs))}
\pgfmathsetmacro{\tA}{\az+\ts}\pgfmathsetmacro{\tB}{\az+180-\ts}
\pgfmathsetmacro{\tg}{tan(\thgs)}
\pgfmathsetmacro{\rtx}{sin(\rktau)*cos(\rkkap)}\pgfmathsetmacro{\rty}{sin(\rktau)*sin(\rkkap)}\pgfmathsetmacro{\rtz}{cos(\rktau)}
\pgfmathsetmacro{\xbx}{cos(\rkkap)*cos(\rktau)*cos(\rkpsi)-sin(\rkkap)*sin(\rkpsi)}
\pgfmathsetmacro{\xby}{sin(\rkkap)*cos(\rktau)*cos(\rkpsi)+cos(\rkkap)*sin(\rkpsi)}
\pgfmathsetmacro{\xbz}{-sin(\rktau)*cos(\rkpsi)}
\pgfmathsetmacro{\ybx}{-cos(\rkkap)*cos(\rktau)*sin(\rkpsi)-sin(\rkkap)*cos(\rkpsi)}
\pgfmathsetmacro{\yby}{-sin(\rkkap)*cos(\rktau)*sin(\rkpsi)+cos(\rkkap)*cos(\rkpsi)}
\pgfmathsetmacro{\ybz}{sin(\rktau)*sin(\rkpsi)}
\begin{scope}[view]
  \coordinate (O) at (0,0,0);
  \coordinate (I) at (-22,28,0);
  \coordinate (C) at ({-\rkoff*cos(\az)+\rkdepth*sin(\az)},{-\rkoff*sin(\az)-\rkdepth*cos(\az)},{-\rkalt});
  \coordinate (Cg) at ({-\rkoff*cos(\az)+\rkdepth*sin(\az)},{-\rkoff*sin(\az)-\rkdepth*cos(\az)},0);
  \coordinate (r3p) at (\rtx,\rty,\rtz);
\end{scope}
\pgfmathanglebetweenpoints{\pgfpointanchor{pL}{center}}{\pgfpointanchor{r3p}{center}}
\pgfmathsetmacro{\rkrot}{\pgfmathresult+90}
\begin{scope}[view, canvas is xy plane at z=0]
  \shade[inner color=gray!6, outer color=gray!30] (0,0) circle (\Rg);
  \begin{scope}
    \clip (0,0) circle (\Rg);
    \draw[gray!45, line width=0.1mm] (-42,-42) grid[step=6] (42,42);
  \end{scope}
  \draw[gray!65, line width=0.2mm] (0,0) circle (\Rg);
  \fill[gray!55] (0,0) circle (5);
  \draw[gray!30!black, line width=0.15mm] (0,0) circle (5);
  \fill[black, opacity=0.45, path fading=shadowfade] (Cg) ellipse (2.2 and 1.4);
\end{scope}
\begin{scope}[view]
  \draw[axis, frI] (I) -- ++(6,0,0);
  \draw[axis, frI] (I) -- ++(0,6,0);
  \draw[axis, frI] (I) -- ++(0,0,6);
\end{scope}
\node[anchor=south east, inner sep=0.5pt] at ($(I)+(-0.6,0.7)$) {$\{I\}$};
\begin{scope}[view]
  \shade[left color=gsedge, middle color=gsmid, right color=gsedge, path fading=gsfade]
    (O) -- ({64*\tg*cos(\tA)},{64*\tg*sin(\tA)},-64) -- ({64*\tg*cos(\tB)},{64*\tg*sin(\tB)},-64) -- cycle;
  \foreach \t in {\tA,\tB}
    \draw[gsline, line width=0.3mm] (O) -- ({62*\tg*cos(\t)},{62*\tg*sin(\t)},-62);
  \foreach \lev in {18}{
    \pgfmathsetmacro{\rr}{\lev*\tg}
    \begin{scope}[canvas is xy plane at z=-\lev]
      \draw[gsline, line width=0.17mm, dash pattern=on 0.9mm off 0.6mm, opacity=0.8]
        (\az-180:\rr) arc[start angle=\az-180, end angle=\az, radius=\rr];
    \end{scope}}
  \pgfmathsetmacro{\rr}{\toplev*\tg}
  \begin{scope}[canvas is xy plane at z=-\toplev]
    \draw[gsline, line width=0.17mm, dash pattern=on 0.9mm off 0.6mm, opacity=0.4]
      (\az-180:\rr) arc[start angle=\az-180, end angle=\az, radius=\rr];
  \end{scope}
  \coordinate (gsE) at ({34*\tg*cos(\tA)},{34*\tg*sin(\tA)},-34);
  \draw[pathc, line width=0.33mm, dash pattern=on 1.6mm off 0.8mm] (O) -- (0,0,-\hzero);
  \fill[pathc] (0,0,-\hzero) circle[radius=0.5mm];
  \draw[pathc, line width=0.33mm, -{Triangle[length=1.9mm, width=1.4mm]}] (0,0,-25) -- (0,0,-22.5);
  \foreach \lev in {18}{
    \pgfmathsetmacro{\rr}{\lev*\tg}
    \begin{scope}[canvas is xy plane at z=-\lev]
      \draw[gsline, line width=0.25mm] (\az:\rr) arc[start angle=\az, end angle=\az+180, radius=\rr];
    \end{scope}}
  \begin{scope}[plane origin={(0,0,0)}, plane x={({cos(\az)},{sin(\az)},0)}, plane y={(0,0,-1)}, canvas is plane]
    \draw[gsline, line width=0.22mm] (90:7) arc[start angle=90, end angle=90-\thgs, radius=7];
    \path (90-0.5*\thgs:9.6) node {$\theta_\mathrm{gs}$};
  \end{scope}
\end{scope}
\fill (pL) circle[radius=0.45mm];
\draw[vec, line width=0.24mm, shorten >=0.5mm] (I) -- (pL) node[pos=0.5, anchor=north, inner sep=1.2pt] {$\mathbf{p}_L$};
\node at ($(pL)+(24,41)$) {$\mathcal{S}$};
\node[anchor=west, inner sep=0.5pt] (clab) at ($(pL)+(29.5,25.5)$) {$c(\mathbf{p}_\delta)=0$};
\draw[gsline, line width=0.22mm, -{Stealth[length=1.1mm, width=0.9mm]}, shorten >=0.4mm] ($(clab.west)+(0,-0.3)$) -- (gsE);
\draw[gray!60!black, line width=0.15mm, dash pattern=on 0.4mm off 0.6mm] (Cg) -- ($(Cg)!0.55!(C)$);
\begin{scope}[transform canvas={shift={(C)}, rotate=\rkrot, scale=\rks}]
  \begin{scope}[scale=2.84527, rotate=20.53, yscale=-1, shift={(-108.9228pt,-137.3393pt)}]
    \fill[flame, path fading=south] svg "M 105.17136,153.22613 c -2.55792,2.78718 -4.56264,6.73351 -5.802061,8.071 0.11564,-2.03237 1.407421,-6.18315 2.073351,-9.45823 z";
    \shadedraw[left color=black!85, right color=black!15, draw=black, line width=0.19mm]
      svg "M 102.48331,149.86142 c -0.45315,0.24709 -0.93895,0.88921 -1.44178,1.73734 l 4.52449,1.6944 c 0.15782,-1.11584 0.10899,-1.70861 -0.0979,-2.31394 z";
    \shadedraw[left color=rkdark, right color=rklight, draw=black, line width=0.25mm]
      svg "M 101.09826,149.27738 l 10.33693,-27.51263 c 1.83653,-3.66688 4.32868,-6.51459 5.68718,-6.02047 1.36209,0.55076 1.06305,4.27674 0.0644,8.27844 l -10.22572,27.43513 z";
  \end{scope}
\end{scope}
\coordinate (Nz) at ($(C)+(\rkrot-90:{16.1*\rks})$);
\draw[vec, thrc, line width=0.3mm] ($(Nz)+(\rkrot-93.4:7)$) -- ($(Nz)+(\rkrot-93.4:0.2)$);
\node[anchor=east, inner sep=0.5pt] at ($(Nz)+(\rkrot-93.4:7)+(-0.4,0.3)$) {$\mathbf{u}$};
\begin{scope}[view, canvas is xy plane at z=-\toplev]
  \pgfmathsetmacro{\rr}{\toplev*\tg}
  \draw[gsline, line width=0.2mm, opacity=0.45] (\az:\rr) arc[start angle=\az, end angle=\az+180, radius=\rr];
\end{scope}
\begin{scope}[view]
  \draw[axis] (C) -- ++(4.6*\xbx,4.6*\xby,4.6*\xbz);
  \draw[axis] (C) -- ++(4.6*\ybx,4.6*\yby,4.6*\ybz);
  \draw[axis] (C) -- ++(5.8*\rtx,5.8*\rty,5.8*\rtz) coordinate (r3tip);
\end{scope}
\node[anchor=north west, inner sep=0.5pt] at ($(r3tip)+(1.2,-0.4)$) {$\mathbf{r}_3$};
\node[anchor=north west, inner sep=0.5pt] at ($(C)+(1.5,-2.3)$) {$\{B\}$};
\fill[white] (C) circle[radius=0.55mm];
\fill[black] (C) -- ++(0.55mm,0) arc[start angle=0, end angle=90, radius=0.55mm] -- cycle;
\fill[black] (C) -- ++(-0.55mm,0) arc[start angle=180, end angle=270, radius=0.55mm] -- cycle;
\draw[line width=0.12mm] (C) circle[radius=0.55mm];
\coordinate (Pd) at ($(C)+(\rkrot+90:\rkdelta)$);
\coordinate (Pdes) at (Pd -| pL);
\draw[gray!70!black, line width=0.17mm, dash pattern=on 0.5mm off 0.5mm] (Pd) -- (Pdes);
\filldraw[fill=white, draw=black, line width=0.17mm] (Pd) circle[radius=0.5mm];
\filldraw[fill=pathc, draw=white, line width=0.12mm] (Pdes) circle[radius=0.6mm];
\node[anchor=west, inner sep=0.5pt] at ($(Pdes)+(1.0,0)$) {$\mathbf{p}_d(\gamma)$};
\draw[line width=0.17mm, {Bar[width=1.1mm]Stealth[length=0.8mm,width=0.7mm]}-{Stealth[length=0.8mm,width=0.7mm]Bar[width=1.1mm]}]
  ($(C)+(\rkrot+180:2.7)$) -- ($(Pd)+(\rkrot+180:2.7)$);
\draw[gray!60!black, line width=0.1mm] ($(C)+(\rkrot+180:0.7)$) -- ($(C)+(\rkrot+180:3.3)$)
  ($(Pd)+(\rkrot+180:0.6)$) -- ($(Pd)+(\rkrot+180:3.3)$);
\node[anchor=east, inner sep=0.5pt] at ($($(C)!0.5!(Pd)$)+(\rkrot+180:3.2)$) {$\delta$};
\draw[line width=0.17mm, {Bar[width=1.1mm]Stealth[length=0.8mm,width=0.7mm]}-{Stealth[length=0.8mm,width=0.7mm]Bar[width=1.1mm]}]
  ($(C)+(\rkrot+180:2.7)$) -- ($(Nz)+(\rkrot+180:2.7)$);
\draw[gray!60!black, line width=0.1mm] ($(Nz)+(\rkrot+180:0.9)$) -- ($(Nz)+(\rkrot+180:3.3)$);
\node[anchor=east, inner sep=0.5pt] at ($($(C)!0.5!(Nz)$)+(\rkrot+180:3.2)$) {$\ell$};
\draw[vec, line width=0.24mm, shorten >=0.55mm] (I) -- (C) node[pos=0.55, anchor=south east, inner sep=1pt] {$\mathbf{p}$};
\coordinate (Ic) at (74,24); \def\Ir{9.6}
\pgfmathsetmacro{\ct}{cot(\thgs)}
\draw[gray!70!black, line width=0.15mm] (pL) circle[radius=2.1mm];
\path let \p1=($(Ic)-(pL)$), \n1={veclen(\x1,\y1)}, \n2={atan2(\y1,\x1)}, \n3={asin((\Ir-2.1)/(\n1/1mm))} in
  \pgfextra{\xdef\figLDang{\n2}\xdef\figLDbet{\n3}};
\foreach \sg in {1,-1}
  \draw[gray!70!black, line width=0.15mm] ($(pL)+(\figLDang+\sg*90+\sg*\figLDbet:2.1)$) -- ($(Ic)+(\figLDang+\sg*90+\sg*\figLDbet:\Ir)$);
\fill[white] (Ic) circle (\Ir);
\begin{scope}
  \clip (Ic) circle (\Ir);
  \begin{scope}[shift={($(Ic)+(0,-1.5)$)}]
    \fill[gsfill, opacity=0.22] plot[domain=-11:11, samples=81] (\x,{\ct*(sqrt(\x*\x+\epsin*\epsin)-\epsin)}) -- (11,20) -- (-11,20) -- cycle;
    \draw[pathc, line width=0.3mm, dash pattern=on 1.6mm off 0.8mm] (0,0) -- (0,20);
    \draw[gsline, line width=0.25mm, dash pattern=on 0.9mm off 0.6mm] (-11,{11*\ct}) -- (0,0) -- (11,{11*\ct});
    \draw[gsline, line width=0.35mm] plot[domain=-11:11, samples=81] (\x,{\ct*(sqrt(\x*\x+\epsin*\epsin)-\epsin)});
    \fill (0,0) circle[radius=0.45mm];
  \end{scope}
\end{scope}
\draw[gray!70!black, line width=0.2mm] (Ic) circle (\Ir);
\end{tikzpicture}
\caption{Landing scenario: reference frames, control point, desired path, and glideslope safe set.}
\label{fig:landing}
\end{figure}

 Consider two reference frames (Fig.~\ref{fig:landing}): an inertial frame, denoted by \{$I$\}, and a body-fixed frame, denoted by \{$B$\}, both right-handed and orthonormal. The third axis of $\{I\}$ points along the local gravity vector at the landing site, and its first two axes span the orthogonal plane. The origin of $\{B\}$ coincides with the center of mass of the vehicle, its third axis is aligned with the longitudinal axis of the vehicle, pointing aft, and its first two axes span the orthogonal plane, following the same convention as the axes of $\{I\}$.
Let $\b{p}, \b{v} \in \mathbb{R}^3$ be the inertial position and velocity of the center of mass, $\b{R}\in \mathrm{SO}(3)$ the rotation matrix from \{$B$\} to \{$I$\}, and $\bs{\omega}\in\mathbb{R}^3$ the angular velocity of $\{B\}$ with respect to $\{I\}$, expressed in $\{B\}$. The mass is $m>0$, $\b{J}\in \mathbb{R}^{3\times3}$ is the positive definite inertia tensor, the thrust-vector input, expressed in \{$B$\}, is denoted by $\b{u} \in \mathbb{R}^3$, $g>0$ is the gravitational acceleration, $\bs{\ell}\in\mathbb{R}^3$ is the position of the point of application of the thrust with respect to the center of mass, expressed in \{$B$\}, and $\b{f}_a \in \mathbb{R}^3$ and $\bs{\tau}_a\in \mathbb{R}^3$ are the aerodynamic force, expressed in \{$I$\}, and moment, expressed in \{$B$\}.

The Newton-Euler equations of motion are
\begin{subequations}\label{eq:kinematics}
	\begin{empheq}[left=\empheqlbrace]{align}
		\dot{\b{p}}&= \b{v}\,,\\[4pt]
				m(t)\dot{\b{v}}
		&= 	m(t)g(t)\b{e}_3 + \b{R}\mathbf{u}	+ \b{f}_a\,, \label{eq:dynamics_a} \\[4pt]
		\dot{\b{R}} &= \mathbf{R}\b{S}(\bs\omega)\,,\\[4pt]
		\b{J}(t)\dot{\bs{\omega}}
		&= -\b{S}(\bs{\omega})\b{J}(t)\bs{\omega}	 + \b{S}(\bs{\ell}(t))\b{u} +\bs{\tau}_a\,, \label{eq:dynamics_b}
 	\end{empheq}
\end{subequations}
where we consider the following assumption.
\begin{assump}\label{ass:time_var}
	The time-varying physical quantities $m(t)$, $\b{J}(t)$, $\bs{\ell}(t)$, and $g(t)$ are taken as known parameters rather than states, i.e., their dynamic influence on the state is neglected.
\end{assump}
Explicit dependence on time is omitted from this point onward, unless otherwise stated.

\subsection{Control-oriented model}
For control design purposes, it is often useful to exploit class-level characteristics of the system and to neglect some dynamic contributions, allowing for the generalization of the methodology and simplifying the design without compromising performance. Under that premise, we abstract the rocket-like vehicle according to Definition~\ref{ass:rocket_like} and adopt the control-oriented model proposed by the authors in \cite{Santos2026}, which requires Assumptions~\ref{ass:spin} and \ref{ass:aero_free}.
\begin{definition}[Rocket-like vehicle]\label{ass:rocket_like}
	A rocket-like vehicle is an idealized axisymmetric rigid body, actuated by a vectored force applied at a point of the longitudinal axis aft of the center of mass. By axial symmetry, the center of mass lies on the longitudinal axis, so that $\bs{\ell} = \ell\,\b{e}_3$, where $\ell>0$ is the distance between the center of mass and the gimbal point of the engine, and the inertia tensor is diagonal, $\b{J} = \text{diag}(j_\perp,j_\perp,j_\parallel)$, where $j_\perp,\,j_\parallel >0$ are the transverse and longitudinal inertia values, respectively.
\end{definition}
\begin{assump}\label{ass:spin}
An auxiliary control system is available to regulate the spinning motion, allowing it to be effectively suppressed, i.e., $\b{e}_3^\mathsf{T}\bs{\omega} \approx 0$.
\end{assump}
\begin{assump}\label{ass:aero_free}
	The aerodynamic force and torque can be neglected for control design, i.e., $\b{f}_a\approx\b{0}$ and $\bs{\tau}_a\approx\b{0}$.
\end{assump}
\begin{remark}\label{rem:assumptions}
 Assumption~\ref{ass:spin} is required for the exact cancellation of the force-torque coupling. Assumption~\ref{ass:aero_free} is reasonable in the terminal landing phase, where the velocity tends to zero, and holds exactly in the absence of an atmosphere; alternatively, aerodynamic models that exploit the axial symmetry of rocket-like vehicles \cite{Pucci2015} can be accommodated in the design. The robustness to both assumptions is assessed in Section~\ref{sec:simulation}.
\end{remark}

The model of \cite{Santos2026} relies on the reduced attitude $\b{r}_3 \in \mathbb{S}^2$, the third column of $\b{R}$, which defines the direction of the longitudinal axis and, by axial symmetry, decouples its motion from the spin about it. The reduced angular velocity $\bs{\Omega}_{12} \coloneq -\b{S}^2(\b{r}_3)\b{R}\,\bs{\omega} \in T_{\b{r}_3}\mathbb{S}^2$ is the projection of the angular velocity, expressed in $\{I\}$, onto the plane orthogonal to $\b{r}_3$, i.e., it retains only the rotation of the longitudinal axis.

The force-torque coupling of thrust vectoring causes non-minimum phase behavior, which is removed by a control point transformation: the control point $\b{p}_\delta \in \mathbb{R}^3$ (see Fig.~\ref{fig:landing}) is placed at the Huygens' center of oscillation \cite{Murray1995},
\begin{equation}\label{eq:p_delta}
	\b{p}_\delta =\b{p} - \delta\b{r}_3 \,,
\end{equation}
where $\delta \coloneq \frac{j_\perp}{m\ell}$ is time-varying.

Under the stated assumptions, the control-oriented model centered on the control point is
\begin{subequations}\label{eq:sys_design}
	\begin{empheq}[left=\empheqlbrace]{align}
		&\dot{\b{p}}_\delta = \b{v}_\delta\,,\\[2pt]
		&\dot{\b{v}}_\delta = g\b{e}_3 + \overline{u_3}\,\b{r}_3\,,\\[2pt]
		&\dot{\b{r}}_3 = -\b{S}(\b{r}_3 )\,\bs{\Omega}_{12}\,,\label{eq:r3_dot_eq}\\[2pt]
		&\dot{\bs{\Omega}}_{12} = \frac{1}{j_\perp}\bs{\Gamma}_u\,,
	\end{empheq}
\end{subequations}
where $\b{v}_\delta\in\mathbb{R}^3$ is the inertial velocity of the control point, $u_3$ is the axial component of the thrust vector $\b{u}$,
\begin{equation}\label{eq:u3_bar_def}
	\overline{u_3}\coloneq \frac{u_3}{m} + \delta\,\bs{\Omega}^\mathsf{T}_{12}\bs{\Omega}_{12}\,,
\end{equation}
and $\bs{\Gamma}_u\in T_{\b{r}_3}\mathbb{S}^2$ is the torque input caused by thrust vectoring, defined as
\begin{equation}\label{eq:Gamma_u}
	\bs\Gamma_u \coloneq  \ell\,\b{S}(\b{r}_3)\b{R}\b{u}\,.
\end{equation}

\subsection{Smooth glideslope constraint}

A typical planetary landing scenario may impose a glideslope constraint to maintain adequate altitude during the approach and prevent the vehicle from colliding with the terrain \cite{Malyuta2022}. This constraint can be formulated as a convex, second-order cone state constraint on the position of the vehicle with respect to the landing site, located at $\b{p}_{L}\in\mathbb{R}^3$, according to
\begin{equation}
	 -\b{e}_3^\mathsf{T}\overbar{\b{p}}_\delta - \cot(\theta_\mathrm{gs})\|{-}\b{S}^2(\b{e}_3)\overbar{\b{p}}_\delta\| \geq 0\,,
\end{equation}
where $\overbar{\b{p}}_\delta\coloneq \b{p}_\delta - \b{p}_{L}$ is the position of the control point with respect to the landing site, and $\theta_\mathrm{gs}\in(0,\pi/2)$ is the glideslope angle, i.e., the half-angle, measured from the vertical, of the cone defining the position constraint. Although convex, the non-smoothness of the cone apex may be problematic when designing a control solution requiring differentiability of the constraint function. To handle this issue, we propose a smooth approximation of the glideslope constraint, which can be expressed through the constraint function $c: \mathbb{R}^3 \to \mathbb{R}$ as
\begin{subequations}\label{eq:smooth_slope}
	\begin{align}
		c(\b{p}_\delta) &\geq 0\,,\\
		c(\b{p}_\delta) &\coloneq -\sin(\theta_\mathrm{gs})\,\b{e}_3^\mathsf{T}\overbar{\b{p}}_\delta\nonumber\\
		&\quad - \cos(\theta_\mathrm{gs})\left[\|{-}\b{S}^2(\b{e}_3)\overbar{\b{p}}_\delta+\epsilon\,\b{e}_3\|-\epsilon\right]\,,
	\end{align}
\end{subequations}
where $\epsilon>0$ is a parameter that ensures global differentiability of the constraint function by smoothing the apex of the cone. The smoothed boundary lies below the original cone by less than $\epsilon\cot(\theta_\mathrm{gs})$, and keeps its apex at the landing site, i.e., $c(\b{p}_L)=0$, so that the landing site lies on the boundary of the constraint. Up to the smoothing, $c(\b{p}_\delta)$ is the left-hand side of the cone constraint multiplied by $\sin(\theta_\mathrm{gs})$, which does not change the boundary and makes $c(\b{p}_\delta)$ the signed distance from $\b{p}_\delta$ to the boundary of the constraint, positive inside it. A graphical representation of the smooth glideslope constraint is shown in Fig.~\ref{fig:landing}, whose inset details the smoothing near the landing site. The smooth glideslope constraint can be used to define a safe set $\mathcal{S}\subset \mathbb{R}^3$ as
\begin{equation}\label{eq:safe_set}
	\mathcal{S} \coloneq \left\{\b{p}_\delta \in \mathbb{R}^3 : c(\b{p}_\delta) \geq 0\right\}\,,
\end{equation}
i.e., the set of all positions of the control point that satisfy the smooth glideslope constraint.

\subsection{Problem statement and solution strategy}\label{sec:problem}

\begin{problem}[Safe autonomous landing]\label{prob:geral}
	Consider the dynamics and kinematics of a thrust-vector controlled rocket-like vehicle given by \eqref{eq:sys_design}, subject to the smooth glideslope constraint in \eqref{eq:smooth_slope}. Given an initial condition, defined by the vector $(\b{p}_\delta(0), \b{v}_\delta(0), \b{r}_3(0), \bs{\Omega}_{12}(0))$ with $\b{p}_\delta(0)$ in the interior of $\mathcal{S}$, design a control law for the thrust-vector input $\b{u}$ that ensures the vehicle reaches a predefined landing site $\b{p}_L$, upright and with near-zero velocity, without exiting the safe set $\mathcal{S}$ in \eqref{eq:safe_set}.
\end{problem}

Given Problem~\ref{prob:geral}, we formulate the landing as a path following problem, in which the control point converges to a desired path $\b{p}_d(\gamma)\in\mathbb{R}^3$ (see Fig.~\ref{fig:landing}), parameterized by $\gamma\in\mathbb{R}$, whose progression $\gamma(t)$ is not fixed in advance and is used to enforce the glideslope constraint. The path lies in the interior of $\mathcal{S}$ for $\gamma<\gamma_f$ and ends at the landing site with a vanishing tangent vector, i.e., $\b{p}_d(\gamma_f)=\b{p}_L$ and $\partial\b{p}_d/\partial\gamma\,(\gamma_f)=\b{0}$ for some $\gamma_f>0$, so that the desired velocity $\dot{\b{p}}_d=(\partial\b{p}_d/\partial\gamma)\,\dot\gamma$ vanishes there for any progression. The control is switched off when $\gamma$ reaches $\gamma_f$ or, in practice, when the vehicle is sufficiently close to the landing site. Problem~\ref{prob:geral} is therefore split into two subproblems: the first determines the thrust-vector input $\b{u}$ for a given progression, and the second determines only the progression.

\begin{subproblem}[Path following]\label{prob:pf}
 Given a sufficiently smooth desired path $\b{p}_d(\gamma)$ and a progression $\gamma(t)$ of the path parameter with bounded time derivatives, design a control law for the thrust-vector input $\b{u}$ that ensures the position of the control point converges to the desired path, i.e., $\b{p}_\delta(t)-\b{p}_d(\gamma(t))\to\b{0}$ as $t\to\infty$.
\end{subproblem}

\begin{subproblem}[Constrained path governor]\label{prob:pg}
 Given a path following control law which solves Problem~\ref{prob:pf}, design a control law for the path parameter progression $\gamma(t)$ that, for any initial condition with $\b{p}_\delta(0)$ in the interior of $\mathcal{S}$, ensures that (i) the vehicle's position remains within the safe set \eqref{eq:safe_set}, i.e., $\b{p}_\delta(t)\in\mathcal{S}$ for all $t\geq0$; and
 (ii) $\gamma(t)$ follows a nominal progression that drives it to $\gamma_f$\ whenever this is compatible with the condition used to enforce (i).
\end{subproblem}

Problem~\ref{prob:pg} is addressed in Section~\ref{sec:generic} for a general class of systems, assuming that a solution to Problem~\ref{prob:pf} is available. Section~\ref{sec:application} then solves Problem~\ref{prob:pf} for the rocket-like vehicle, shows that the resulting closed loop belongs to this class, and combines both results to address Problem~\ref{prob:geral}.

\section{CBF-governed Path Following }\label{sec:generic}
We wish to exploit the degree of freedom introduced by the path following formulation, i.e., the progression of the path parameter, to enforce safety requirements formulated as state constraints. To that end, we make use of CBFs. By considering a pre-stabilized system and only adjusting the path progression through a CBF-based path governor, which acts as a safety filter on the path progression, we aim at retaining the stability properties while ensuring safety. We start by considering the dynamic system
\begin{subequations}\label{eq:gen_sys}
\begin{equation}
	\dot{\b{x}} = f(\b{x},\b{u})\,,
\end{equation}
\begin{equation}
	\b{y} = h(\b{x})\,,
\end{equation}
\end{subequations}
where $f:\mathbb{R}^n\times\mathbb{R}^m \to \mathbb{R}^n$ and $h:\mathbb{R}^n \to \mathbb{R}^p$ are locally Lipschitz continuous functions, $\b{x}\in\mathbb{R}^n$ is the system state, $\b{y}\in\mathbb{R}^p$ is the output, and $\b{u}\in\mathbb{R}^m$ is the control input.
\begin{assump}\label{ass:pre_stab}
Given a sufficiently smooth parameterized reference $\b{y}_d(\gamma) \in \mathbb{R}^p$, where $\gamma(t)$ is the path parameter, there is a control law
\begin{equation}\label{eq:pre_law_1}
	\b{u} = \kappa\left(\b{x},\b{y}_d,\b{y}_d^{(1)},\ldots,\b{y}_d^{(r)}\right)\,,
\end{equation}
where $r$ denotes the highest order of the time derivatives of $\b{y}_d$ used by the law, which ensures asymptotic convergence of $\b{y}$ to $\b{y}_d(\gamma)$ for any initial condition $\b{x}(0)$ belonging to a given set $\mathcal{I}$ and any $\gamma$ of class $\mathcal{C}^r$ for which $t\mapsto\b{y}_d(\gamma)$ has bounded derivatives up to order $r$.
\end{assump}
If the law in \eqref{eq:pre_law_1} is dynamic, as the one designed in Section~\ref{sec:application}, which includes integral action, its states are appended to $\b{x}$, so that \eqref{eq:gen_sys} describes the plant together with the controller. By defining the vector $\bs{\gamma} \coloneq \left[\gamma\:\:\gamma^{(1)}\:\:\ldots\:\:\gamma^{(r-1)}\right]^\mathsf{T}$, and noting that, by the chain rule, $\b{y}_d^{(k)}$ depends only on $\gamma,\ldots,\gamma^{(k)}$, the control law in \eqref{eq:pre_law_1} can be expressed as
\begin{equation}\label{eq:pre_law_2}
	\b{u} = \kappa\left(\b{x},\bs{\gamma},\nu\right)\,,
\end{equation}
where $\nu\coloneq\gamma^{(r)}$. Using \eqref{eq:chain_integrators} and \eqref{eq:pre_law_2}, the pre-stabilized system can be written as
\begin{subequations}\label{eq:pre_stab_dyn}
\begin{equation}
	\dot{\b{x}} = f(\b{x},\kappa\left(\b{x},\bs{\gamma},\nu\right))\coloneq f_s(\b{x},\bs{\gamma},\nu)\,,\label{eq:pre_stab_dyn_x}
\end{equation}
\begin{equation}
    \dot{\bs{\gamma}} = \b{A}_r\bs{\gamma}+\b{B}_r\nu\,. \label{eq:path_dynamics}
\end{equation}
\end{subequations}
\begin{assump}\label{ass:affine}
	The pre-stabilized system in \eqref{eq:pre_stab_dyn_x} is input affine, i.e., it can be written in the form
	\begin{equation}
		\dot{\b{x}} = f_s(\b{x},\bs{\gamma}) + g_s(\b{x},\bs{\gamma})\nu\,.
	\end{equation}
\end{assump}
By defining a new combined state as $\b{z}\coloneq \left[\,\b{x}^\mathsf{T}\:\:\bs{\gamma}^\mathsf{T}\,\right]^\mathsf{T}$, with $\b{z} \in \mathbb{R}^{n+r}$, the pre-stabilized dynamics in \eqref{eq:pre_stab_dyn} can be expressed in the compact form
\begin{subequations}\label{eq:final_stab_sys}
\begin{equation}
	\dot{\b{z}} = f_z(\b{z}) + g_z(\b{z})\nu\,,
\end{equation}
\begin{equation}
	f_z\coloneq  \begin{bmatrix}
		f_s(\b{x},\bs{\gamma})\\
		\b{A}_r\bs{\gamma}
	\end{bmatrix}\,,\hspace{10pt}
	g_z\coloneq
	\begin{bmatrix}
		g_s(\b{x},\bs{\gamma})\\
		\b{B}_r
	\end{bmatrix}\,.
\end{equation}
\end{subequations}
The input $\nu$ determines, through \eqref{eq:path_dynamics}, the reference applied to the pre-stabilized system, and can therefore be used to enforce state constraints. Since $\gamma,\ldots,\gamma^{(r-1)}$ are states of \eqref{eq:path_dynamics}, modifying only $\nu$ leaves them continuous, so that the progression remains as many times differentiable as required by \eqref{eq:pre_law_2}. Since the relative degree of a state constraint with respect to $\nu$ is in general larger than one, we rely on ECBFs \cite{Ames2019}. Consider a safe set
\begin{equation}\label{eq:safe_set_z}
	\mathcal{S}_z \coloneq\left\{\b{z}\in\mathbb{R}^{n+r}:B(\b{z})\geq0\right\},
\end{equation}
defined by a function $B:\mathbb{R}^{n+r}\to\mathbb{R}$. The Lie derivatives of $B$ along the vector fields of \eqref{eq:final_stab_sys} are denoted by $L_fB(\b{z})\coloneq\frac{\partial B}{\partial\b{z}}f_z(\b{z})$ and $L_gB(\b{z})\coloneq\frac{\partial B}{\partial\b{z}}g_z(\b{z})$, with $L_f^{0}B\coloneq B$ and $L_f^{i}B\coloneq L_f(L_f^{i-1}B)$, where the subscript $z$ is omitted for brevity. We assume that $B$ satisfies the following assumption.
\begin{assump}\label{ass:ecbf}
	The function $B$ is of class $\mathcal{C}^\rho$ and the control $\nu$ first appears explicitly in its $\rho$-th time derivative, i.e., $B$ has relative degree $\rho$ with respect to $\nu$: $L_g L_f^{i} B(\b{z}) = 0$ for $i\in\{0,\ldots,\rho-2\}$ and $L_g L_f^{\rho-1} B(\b{z})\neq0$, for all $\b{z} \in \mathcal{S}_z$.
\end{assump}
Under Assumption~\ref{ass:ecbf}, $B^{(\rho)}=\mu$ on $\mathcal{S}_z$, with the virtual input $\mu\coloneq L_f^{\rho} B(\b{z}) + L_g L_f^{\rho-1} B(\b{z})\,\nu$, so that $\bs{\eta}_b(\b{z}) \coloneq \left[\,B(\b{z})\:\:B^{(1)}(\b{z})\:\:\dots\:\:B^{(\rho-1)}(\b{z})\,\right]^\mathsf{T}\in\mathbb{R}^{\rho}$ evolves according to
\begin{equation}\label{eq:eta_b_dyn}
	\dot{\bs{\eta}}_b(\b{z}) = \b{A}_\rho\bs{\eta}_b(\b{z}) + \b{B}_\rho\mu\,,\hspace{10pt}
	B(\b{z}) = \b{C}\bs{\eta}_b(\b{z})\,,
\end{equation}
where $\b{C}\coloneq\left[\,1\:\:0\:\:\dots\:\:0\,\right] \in \mathbb{R}^{1\times\rho}$.
\begin{definition}[Exponential CBF \cite{Ames2019}]\label{def:ecbf}
	The function $B$ is an ECBF for system \eqref{eq:final_stab_sys} if there exists a vector $\b{k}_\alpha\in\mathbb{R}^\rho$ such that
\begin{equation}\label{eq:ecbf_cond}
	\sup_{\nu \in \mathbb{R}} \left[ L_f^{\rho} B(\b{z}) + L_g L_f^{\rho-1} B(\b{z})\,\nu \right] \geq -\b{k}_\alpha^\mathsf{T} \bs{\eta}_b(\b{z})
\end{equation}
for all $\b{z} \in \mathcal{S}_z$.
\end{definition}
\begin{lemma}[\cite{Ames2019,Nguyen2016}]\label{th:ecbf}
Let $B$ be an ECBF with a vector $\b{k}_\alpha$ such that $\b{A}_\rho-\b{B}_\rho\b{k}_\alpha^\mathsf{T}$ is Hurwitz and total negative, i.e., with real, strictly negative eigenvalues $\lambda_i$, $i\in\{1,\ldots,\rho\}$, such that $-\lambda_i\geq -\dot{v}_{i-1}(\b{z}(0))/v_{i-1}(\b{z}(0))$ for $i\in\{1,\ldots,\rho-1\}$, where
\begin{equation}\label{eq:v_family}
	v_0(\b{z})\coloneq B(\b{z})\,,\hspace{10pt}v_i(\b{z})\coloneq \dot{v}_{i-1}(\b{z}) - \lambda_i v_{i-1}(\b{z})\,.
\end{equation}
If $B(\b{z}(0))\geq0$ and $\mu\geq-\b{k}_\alpha^\mathsf{T}\bs{\eta}_b(\b{z})$ along the solution, then $B(\b{z}(t)) \geq \b{C}e^{(\b{A}_\rho-\b{B}_\rho\b{k}_\alpha^\mathsf{T})t}\bs{\eta}_b(\b{z}(0))\geq0$ for all $t\geq0$, i.e., $\mathcal{S}_z$ is forward invariant.
\end{lemma}
\begin{pf}
	A complete proof can be found in \cite{Ames2019,Nguyen2016}; the condition for $i=\rho$ is not needed, since $v_\rho=\mu+\b{k}_\alpha^\mathsf{T}\bs\eta_b(\b{z})$ is the ECBF constraint itself \cite[Remark~6]{Nguyen2016}.
\end{pf}

\subsection{ECBF-QP path governor}\label{sec:safety_filter}

Given an ECBF, we can incorporate inequality \eqref{eq:ecbf_cond} as a constraint into an optimization problem to determine the input $\nu$ that ensures satisfaction of the ECBF condition while minimizing the deviation from a nominal path progression. This optimization problem can be formulated as a QP as follows:
\begin{equation}\label{eq:ecbf_qp}
\begin{aligned}
    \nu = \operatorname*{argmin}_{\nu \in \mathbb{R}} \quad & \frac{1}{2}\|\nu - \nu_\mathrm{nom}\|^2 \hspace{20pt} \text{(ECBF-QP)}\\
    \text{s.t.} \quad & \mu \geq -\b{k}_\alpha^\mathsf{T} \bs{\eta}_b(\b{z}) \\
\end{aligned}
\end{equation}
where $\nu_\mathrm{nom}$ is the nominal value of $\nu$, which can be designed based on the desired speed profile along the path. The constraint ensures that the ECBF condition is satisfied, thereby guaranteeing that the system remains within the safe set $\mathcal{S}_z$ in \eqref{eq:safe_set_z}, specified by the selected ECBF. The ECBF-QP in \eqref{eq:ecbf_qp} has a single constraint which is affine in the scalar input $\nu$, therefore it admits the closed-form solution \cite{Xu2015}
\begin{equation}\label{eq:ecbf_qp_sol}
\nu^{*} =
\begin{cases}
\nu_{\mathrm{nom}},  &\text{if } \mu|_{\nu=\nu_\mathrm{nom}} \ge -\b{k}_{\alpha}^\mathsf{T}\bs{\eta}_{b}(\mathbf{z}) \\[8pt]
\dfrac{-\b{k}_{\alpha}^\mathsf{T}\bs{\eta}_{b}(\mathbf{z}) - L_f^{\rho}B(\mathbf{z})}{L_gL_f^{\rho-1}B(\mathbf{z})},  &\text{otherwise}
\end{cases}
\end{equation}
Here, $\mu|_{\nu=\nu_\mathrm{nom}}$ denotes $\mu$ evaluated at $\nu=\nu_\mathrm{nom}$. The governor is said to be \textit{inactive} at time $t$ if $\nu_\mathrm{nom}$ satisfies the constraint in \eqref{eq:ecbf_qp}, so that $\nu=\nu_\mathrm{nom}$ by \eqref{eq:ecbf_qp_sol}, and \textit{active} otherwise, in which case the constraint holds with equality, $\mu=-\b{k}_\alpha^\mathsf{T}\bs\eta_b(\b{z})$.

\begin{remark}\label{rem:feasibility}
	The ECBF-QP path governor in \eqref{eq:ecbf_qp} is a first approach to exploit the path following degree of freedom when safety constraints can be formulated as a single ECBF of the variable $\b{z}$. Feasibility of the QP relies on $L_g L_f^{\rho-1} B(\b{z})\neq0$ in Assumption~\ref{ass:ecbf}, which is not guaranteed in general and depends on the choice of the ECBF, the path to be followed, and the pre-stabilizing law. Other constraints can be included in the optimization problem, for instance through additional ECBFs, or by imposing bounds on $\nu$.
\end{remark}

Considering the path following task, we wish to impose the nominal dynamics $\dot{\gamma} = v_d(t)$, which determine the evolution of the path parameter according to a desired speed profile, denoted by $v_d(t) \in \mathbb{R}$. To achieve that goal, a nominal control law can be set as
\begin{subequations}\label{eq:nom_path_law}
\begin{equation}
	\nu_\mathrm{nom} \coloneq -\b{k}_{\gamma}^\mathsf{T} \left(\overbar{\bs\gamma}-\overbar{\bs\gamma}_\mathrm{nom}\right) + v_d^{(r-1)}\,,
\end{equation}
	\begin{equation}
		\overbar{\bs\gamma}\coloneq\left[\,\gamma^{(1)}\:\:...\:\:\gamma^{(r-1)}\,\right]^\mathsf{T}\,,
	\end{equation}
	\begin{equation}
		\overbar{\bs\gamma}_\mathrm{nom}\coloneq\left[\,v_d\:\:...\:\:v_d^{(r-2)}\,\right]^\mathsf{T}\,,
	\end{equation}
\end{subequations}
where $\b{k}_\gamma\in{\mathbb{R}^{r-1}}$ is a constant gain vector.
\begin{lemma}\label{lemma:nominal_convergence}
	For $\b{k}_\gamma$ such that $\b{A}_{(r-1)}-\b{B}_{(r-1)}\b{k}_\gamma^\mathsf{T}$ is Hurwitz, if $\nu=\nu_\mathrm{nom}$, with $\nu_\mathrm{nom}$ in \eqref{eq:nom_path_law}, on an interval $[T,\infty)$, $T\geq0$, then $\gamma^{(i)}-v_d^{(i-1)}$ converges exponentially to zero for $i\in\{1,\ldots,r-1\}$. In particular, $\dot\gamma$ converges exponentially to the desired speed profile $v_d$.
\end{lemma}

\begin{pf}
	Since $\overbar{\bs\gamma}$ collects the last $r-1$ entries of $\bs\gamma$, \eqref{eq:path_dynamics} gives $\dot{\overbar{\bs\gamma}}=\b{A}_{(r-1)}\overbar{\bs\gamma}+\b{B}_{(r-1)}\nu$, while $\dot{\overbar{\bs\gamma}}_\mathrm{nom}=\b{A}_{(r-1)}\overbar{\bs\gamma}_\mathrm{nom}+\b{B}_{(r-1)}v_d^{(r-1)}$ by definition. Hence, with $\nu=\nu_\mathrm{nom}$ in \eqref{eq:nom_path_law}, the error $\bs e_\gamma\coloneq\overbar{\bs\gamma}-\overbar{\bs\gamma}_\mathrm{nom}$, whose entries are $\gamma^{(i)}-v_d^{(i-1)}$, satisfies $\dot{\bs e}_\gamma=(\b{A}_{(r-1)}-\b{B}_{(r-1)}\b{k}_\gamma^\mathsf{T})\bs e_\gamma$ on $[T,\infty)$, and converges exponentially to zero, since this matrix is Hurwitz \cite[Th.~8.1]{Hespanha2009}.
\end{pf}

\begin{theorem}\label{thm:generic}
	Under Assumptions~\ref{ass:pre_stab}--\ref{ass:ecbf}, and provided that the conditions on $\b{k}_\alpha$ in Lemma~\ref{th:ecbf} are satisfied, for any initial condition such that $\b{x}(0)\in\mathcal{I}$ and $B(\b{z}(0))>0$, the combination of the pre-stabilizing control law in \eqref{eq:pre_law_2} with the ECBF-QP path governor in \eqref{eq:ecbf_qp} ensures that the safe set $\mathcal{S}_z$ in \eqref{eq:safe_set_z} is forward invariant{{. If, in addition, $t\mapsto\b{y}_d(\gamma(t))$ has bounded derivatives up to order $r$ along the resulting solution, the output $\b{y}$ of the system \eqref{eq:gen_sys} asymptotically tracks the desired path $\b{y}_d(\gamma)$}. Moreover, if the governor is inactive on an interval $[T,\infty)$, $T\geq0$, $\dot\gamma$ converges exponentially to $v_d$.}
\end{theorem}
\begin{pf}
	By Assumption~\ref{ass:ecbf}, $L_g L_f^{\rho-1} B(\b{z})\neq0$ in $\mathcal{S}_z$, so the ECBF-QP is feasible and its solution is given in closed form. The ECBF-QP path governor modifies the $r$-th order derivative $\nu$ of the path parameter $\gamma$ to ensure that the system remains within the safe set $\mathcal{S}_z$ in \eqref{eq:safe_set_z} defined by the ECBF. By Lemma~\ref{th:ecbf}, if the conditions on $\b{k}_\alpha$ are satisfied, then the ECBF condition guarantees that $B(\b{z}(t)) \geq 0$ for all $t \geq 0$, provided that $B(\b{z}(0)) \geq 0$. This implies that the system state $\b{z}(t)$ remains within the safe set $\mathcal{S}_z$ for all time. Since the solution of the ECBF-QP is continuous, the resulting progression $\gamma(t)$ is of class $\mathcal{C}^r$, so that, under the additional boundedness condition, Assumption~\ref{ass:pre_stab} ensures that $\b{y}$ asymptotically tracks $\b{y}_d(\gamma)$ for any $\b{x}(0) \in \mathcal{I}$. Finally, if the governor is inactive on $[T,\infty)$, then $\nu=\nu_\mathrm{nom}$ on that interval, and Lemma~\ref{lemma:nominal_convergence} gives the exponential convergence of $\dot\gamma$ to $v_d$.
\end{pf}

\section{Application to Autonomous Rocket Landing}\label{sec:application}
Leveraging the CBF-governed path following framework{\ of Section~\ref{sec:generic}}, we now focus on the application of the proposed approach to the autonomous landing of rocket-like vehicles, thereby addressing Problem~\ref{prob:geral}. We start by matching this framework to the specific dynamics of the rocket-like vehicle described in the control-oriented model \eqref{eq:sys_design}. With the thrust vector $\b{u}$ as input, the state and output vectors are defined as
\begin{equation}\label{eq:state_input_output}
	\b{x} \coloneq \left[\,\b{p}_\delta^\mathsf{T}\:\:\b{v}_\delta^\mathsf{T}\:\:\b{r}_3^\mathsf{T}\:\:\bs{\Omega}_{12}^\mathsf{T}\,\right]^\mathsf{T}\,,\hspace{10pt}
	\b{y} \coloneq \b{p}_\delta\,,
\end{equation}
so that \eqref{eq:sys_design} is in the form of \eqref{eq:gen_sys}.

\subsection{Nominal path generation}

Since the glideslope constraint is enforced by the path governor, the nominal path only needs to lie inside the glideslope cone and to satisfy the boundary conditions at the landing site. We adopt the simplest such path, which requires no trajectory computation: the vertical line through the landing site, traversed with a constant deceleration from the initial vertical velocity to zero (a constant deceleration resembles the maximum-thrust arc of the bang-bang thrust profiles that characterize propellant-optimal descents \cite{Lu2023}). The path thus starts at the initial altitude, and any horizontal offset is left to the path following law. If the offset is not removed as fast as the nominal descent proceeds, the vehicle approaches the boundary of $\mathcal{S}$, and the governor then slows the descent until the offset is reduced. Mathematically, this can be achieved by setting the nominal path progression rate as $\dot{\gamma}_{\mathrm{nom}}\coloneq 1$, meaning that the path parameter $\gamma$ nominally corresponds to the time variable $t$, and defining a nominal altitude path $h_\mathrm{d}:[0,\gamma_f] \to \mathbb{R}$ as
\begin{equation}
	h_\mathrm{d}(\gamma) \coloneq h_0+ \dot{h}_0\gamma +\frac{{\dot{h}_0}^2}{4(h_0-h_L)}\gamma^2\,,
\end{equation}
where $h_0\coloneq-\b{e}_3^\mathsf{T}\b{p}_\delta(0)$ is the initial altitude, $\dot{h}_0\coloneq-\b{e}_3^\mathsf{T}\b{v}_\delta(0)<0$ is the initial vertical velocity, $h_L\coloneq -\b{e}_3^\mathsf{T}\b{p}_L$ is the landing site altitude, and $\gamma_f\coloneq2(h_0-h_L)/|\dot{h}_0|$. The nominal path expressed in $\{I\}$ is then given by
\begin{equation}\label{eq:nominal_path}
	\b{p}_d(\gamma) \coloneq -\b{S}^2(\b{e}_3)\b{p}_{L} - h_\mathrm{d}(\gamma)\b{e}_3\,.
\end{equation}
The path parameter is initialized with $\gamma(0)=0$, $\dot\gamma(0)=1$ and $\ddot\gamma(0)=\gamma^{(3)}(0)=0$, so that the desired altitude and vertical velocity match those of the control point at the initial time.

\subsection{Nonlinear saturated path following control}

The next step is to solve Problem~\ref{prob:pf}, i.e., to design a thrust-vector law $\b{u}$ that drives $\b{p}_\delta$ to the desired path $\b{p}_d(\gamma)$, which pre-stabilizes the system as required by Assumption~\ref{ass:pre_stab}. Given the potentially large initial horizontal displacement, we adopt an implicitly saturated Lyapunov design \cite{Reis2023}, in which the saturation function \eqref{eq:saturation} limits the translational errors used in the control law, preventing undesirable peaks in the system response and actuation signals.

Consider the position and velocity tracking errors
\begin{equation}
	\b{e}_{\b{p}} \coloneq \b{p}_\delta-\b{p}_d(\gamma)\,,\qquad \b{e}_{\b{v}} \coloneq \b{v}_\delta-\dot{\b{p}}_d(\gamma)=\dot{\b{e}}_{\b{p}}\,,
\end{equation}
and the auxiliary composite error $\b{e}_c\coloneq k_1(\b{e}_{\b{p}}+\beta\,\b{e}_{\b{v}})$, where $k_1,\beta>0$. For the candidate Lyapunov function $V_1 \coloneq \frac{s_1}{k_1}\big(\sqrt{\|\b{e}_c\|^2+s^2_1}-s_1\big) + \frac{1}{2}\|\b{e}_{\b{v}}\|^2$, with $s_1>0$, \eqref{eq:saturation} gives
\begin{equation}\label{eq:V1_dot_1}
	\dot{V}_1 = \bs{\sigma}^\mathsf{T}(\b{e}_c,s_1)\,\b{e}_{\b{v}} + \big(\beta\bs\sigma(\b{e}_c,s_1)+\b{e}_{\b{v}}\big)^\mathsf{T}\dot{\b{e}}_{\b{v}}\,.
\end{equation}
Following the structure of \eqref{eq:V1_dot_1}, we introduce an integral error $\b{e}_I\in\mathbb{R}^3$ with dynamics $\dot{\b{e}}_I\coloneq(\partial V_1/\partial\b{e}_{\b{v}})^\mathsf{T}=\beta\bs\sigma(\b{e}_c,s_1)+\b{e}_{\b{v}}$, and augment the candidate Lyapunov function as $V_2 \coloneq V_1 + k_I\,s_2\big(\sqrt{\|\b{e}_I\|^2+s^2_2}-s_2\big)$, with $k_I,s_2>0$. Using $\b{e}_{\b{v}}=\dot{\b{e}}_I-\beta\bs\sigma(\b{e}_c,s_1)$, adding and subtracting $k_2\dot{\b{e}}_I^\mathsf{T}\bs\sigma(\dot{\b{e}}_I,s_3)$, with $k_2,s_3>0$, and expanding $\dot{\b{e}}_{\b{v}}$ with \eqref{eq:sys_design}, we obtain
\begin{equation}\label{eq:v2_dot}
	\dot{V}_2 = -W_1 + \dot{\b{e}}_I^\mathsf{T}\big(\dot{\b{e}}_{\b{v}} + \bs{\zeta}\big) = -W_1 + \dot{\b{e}}_I^\mathsf{T}\big(\overline{u_3}\,\b{r}_3 + \bs{\xi}\big)\,,
\end{equation}
where $W_1\coloneq \beta\|\bs{\sigma}(\b{e}_c,s_1)\|^2 + k_2\dot{\b{e}}_I^\mathsf{T}\bs{\sigma}(\dot{\b{e}}_I,s_3)\geq0$, the vector $\bs{\zeta}\coloneq \bs{\sigma}(\b{e}_c,s_1) + k_I\bs{\sigma}(\b{e}_I,s_2)+  k_2\bs{\sigma}(\dot{\b{e}}_I,s_3)$ collects saturated terms only, and $\bs{\xi} \coloneq g\b{e}_3 - \ddot{\b{p}}_d(\gamma)+ \bs{\zeta}$ is available for feedback. Due to underactuation, the last term in \eqref{eq:v2_dot} cannot be canceled directly, as nothing imposes alignment between $\bs{\xi}$ and $\b{r}_3$. We therefore define a desired direction for the longitudinal axis and impose the axial input
\begin{equation}\label{eq:r3d}
	\b{r}_{3d} \coloneq \bs{\xi}/\|\bs{\xi}\|\,,\qquad \overline{u_3} \coloneq -\bs{\xi}^{\mathsf{T}}\b{r}_{3}\,,
\end{equation}
which are well defined under the following assumption.
\begin{assump}\label{ass:xi}
	Along the closed-loop solutions, there is $\xi_{\min}>0$ such that $\|\bs\xi(t)\|\geq\xi_{\min}$ for all $t\geq0$.
\end{assump}
Since $-\bs\xi$ is the requested thrust acceleration, Assumption~\ref{ass:xi} excludes commanding free fall; it holds, for instance, if $g-s_1-k_Is_2-k_2s_3>\sup_{t\geq0}\b{e}_3^\mathsf{T}\ddot{\b{p}}_d(\gamma(t))$. Substituting \eqref{eq:r3d} into \eqref{eq:v2_dot} yields
\begin{equation}\label{eq:v2_dot_2}
	\dot{V}_2 = -W_1 + \Delta\,,\qquad \Delta \coloneq -\|\bs{\xi}\|\,\dot{\b{e}}_I^\mathsf{T}\b{S}^2(\b{r}_3)\b{r}_{3d}\,.
\end{equation}

It remains to design the torque input so that $\b{r}_3$ aligns with $\b{r}_{3d}$ and cancels $\Delta$. A structurally identical coupling term was addressed in the trajectory tracking design of \cite{Santos2026}, whose attitude loop we adapt here. With the reduced-attitude and reduced angular velocity errors
\begin{equation}
	\b{e_r} \coloneq \b{S}(\b{r}_{3d})\b{r}_{3}\,,\qquad \b{e}_{\bs{\Omega}} \coloneq \bs{\Omega}_{12} - \bs{\Omega}_{12c}\,,
\end{equation}
where $\b{e_r}\in T_{\b{r}_{3}}\mathbb{S}^2 \cap  T_{\b{r}_{3d}}\mathbb{S}^2$ respects the geometry of $\mathbb{S}^2$, the angular velocity command $\bs{\Omega}_{12c}\in T_{\b{r}_3}\mathbb{S}^2$ and the torque command are set as
\begin{align}
	\bs{\Omega}_{12c} &\coloneq -\frac{k_r}{h_r}\b{e_r} + \frac{\|\bs{\xi}\|}{h_r}\b{S}(\b{r}_3)\dot{\b{e}}_I - \b{S}^2(\b{r}_3)\bs{\Omega}_{12d}\,,\\
	\bs{\Gamma}_u &\coloneq -j_\perp\Big(\frac{k_\Omega}{h_\Omega}\b{e}_{\bs{\Omega}} + \frac{h_r}{h_\Omega}\b{e_r} + \b{S}^2(\b{r}_3)\dot{\bs{\Omega}}_{12c}\Big)\,,\label{eq:ang_accel_law}
\end{align}
where $\bs{\Omega}_{12d} \coloneq \b{S}(\b{r}_{3d})\dot{\b{r}}_{3d}\in T_{\b{r}_{3d}}\mathbb{S}^2$ and $k_r,h_r,k_\Omega,h_\Omega>0$. By \eqref{eq:u3_bar_def} and \eqref{eq:Gamma_u}, the thrust-vector input of \eqref{eq:sys_design} is then
\begin{equation}\label{eq:control_law}
	\b{u} \coloneq -\frac{1}{\ell}\b{S}(\b{e}_3)\b{R}^\mathsf{T}\bs{\Gamma}_u-m\left(\bs{\xi}^\mathsf{T}\b{r}_3 + \delta\,\|\bs{\Omega}_{12}\|^2\right)\b{e}_3\,.
\end{equation}
\begin{prop}\label{prop:tracking}
	Consider the control-oriented model \eqref{eq:sys_design} in closed loop with the control law \eqref{eq:control_law}, and suppose that $t\mapsto\b{p}_d(\gamma(t))$ is of class $\mathcal{C}^4$ with bounded derivatives up to order three, and that Assumption~\ref{ass:xi} holds. Then, for every initial condition, $\b{e}_{\b{p}},\b{e}_{\b{v}},\b{e}_I,\b{e}_{\bs\Omega}\to\b{0}$ and $\b{r}_{3d}^\mathsf{T}\b{r}_3\to\pm1$.
\end{prop}

\begin{pf}
	See Appendix~\ref{app:prop1}.
\end{pf}

The ambiguity $\b{r}_3\to\pm\b{r}_{3d}$ is due to the topology of $\mathbb{S}^2$, as no continuous time-invariant feedback globally asymptotically stabilizes $\b{r}_3=\b{r}_{3d}$ \cite{Bhat2000}. Since the initial conditions converging to $-\b{r}_{3d}$ form a set of zero measure, $\b{r}_3\to\b{r}_{3d}$ for almost every initial condition.

\subsection{CBF-based path governor for a glideslope constraint}

Having defined the pre-stabilizing control law in \eqref{eq:control_law}, we can now proceed to the design of a CBF-based path governor to enforce the glideslope constraint in \eqref{eq:smooth_slope}, thereby addressing Problem~\ref{prob:pg}.

\begin{lemma}\label{lemma:closed_loop}
The pre-stabilized system, under the control law in \eqref{eq:control_law}, can be expressed in the form of \eqref{eq:final_stab_sys} with $\b{z} \coloneq  \left[\,\b{x}^\mathsf{T}\:\:\b{e}_I^\mathsf{T}\:\:\bs{\gamma}^\mathsf{T}\,\right]^\mathsf{T}\in\mathbb{R}^{19}$, $\b{x}$ as defined in \eqref{eq:state_input_output}, the integral error $\b{e}_I$ of the control law appended to it as in Section~\ref{sec:generic}, $\bs{\gamma}\coloneq\left[\,\gamma\:\:\dot{\gamma}\:\:\gamma^{(2)}\:\:\gamma^{(3)}\,\right]^\mathsf{T}$, and $\nu\coloneq\gamma^{(4)}$, i.e., $r=4$.
\end{lemma}
\begin{pf}Since $\b{r}_{3d}$ depends on $\ddot{\b{p}}_d(\gamma)$, the highest derivative of $\gamma$ required by the control law is $\gamma^{(4)}$, which appears through $\ddot{\b{r}}_{3d}$ in $\dot{\bs\Omega}_{12c}$. Hence, substituting the control law in \eqref{eq:control_law} and the desired path in \eqref{eq:nominal_path} into \eqref{eq:sys_design} yields a pre-stabilized system in the form of \eqref{eq:final_stab_sys}, where
\begin{equation}
	f_z \coloneq \begin{bmatrix}
		\b{v}_\delta\\
		g\b{e}_3 -\b{r}_3\b{r}_3^\mathsf{T}\bs{\xi} \\
		-\b{S}(\b{r}_3)\bs{\Omega}_{12}\\
		\frac{1}{j_\perp}\overbar{\bs{\Gamma}}_u\\
		\beta\bs\sigma(\b{e}_c,s_1)+\b{e}_{\b{v}}\\
		\b{A}_4\bs{\gamma}
	\end{bmatrix}\,,\hspace{4pt}
	g_z \coloneq \begin{bmatrix}
		\b{0}\\
		\b{0}\\
		\b{0}\\
		\b{g}_{\bs\Omega}\\
		\b{0}\\
		\b{B}_4
	\end{bmatrix}\,,
\end{equation}
where
\begin{equation}\label{eq:g_Omega}
	\b{g}_{\bs\Omega}\coloneq-\frac{1}{\|\bs\xi\|}\frac{\partial h_\mathrm{d}}{\partial\gamma}\,\b{S}^2(\b{r}_3)\b{S}(\b{r}_{3d})\b{e}_3\,,
\end{equation}
and $\overbar{\bs\Gamma}_u$ is the part of the torque command $\bs\Gamma_u$ in \eqref{eq:ang_accel_law} that does not depend on $\nu$, i.e., $\bs\Gamma_u=\overbar{\bs\Gamma}_u+j_\perp\b{g}_{\bs\Omega}\,\nu$.
\end{pf}

With the pre-stabilized system in the form of \eqref{eq:final_stab_sys}, the glideslope constraint is enforced by taking the constraint function itself as the candidate ECBF,
\begin{equation}\label{eq:glide_ecbf}
	B(\b{z}) \coloneq c(\b{p}_\delta)\,,
\end{equation}
so that the set $\mathcal{S}_z$ in \eqref{eq:safe_set_z} is the safe set $\mathcal{S}$ in \eqref{eq:safe_set} lifted to the closed-loop state space, i.e., $\mathcal{S}_z = \left\{\b{z}\in\mathbb{R}^{19} : \b{p}_\delta\in\mathcal{S}\right\}$, and $B$ is{, up to the smoothing,} the signed distance of the control point to the boundary of $\mathcal{S}$. Since $B$ depends on $\b{z}$ only through $\b{p}_\delta$, its relative degree with respect to $\nu$ is determined by the order at which $\nu$ first affects $\b{p}_\delta$.

\begin{lemma}\label{lemma:ecbf_is}
	Consider the pre-stabilized system of Lemma~\ref{lemma:closed_loop} under Assumption~\ref{ass:xi}, and the function $B(\b{z})$ in \eqref{eq:glide_ecbf}. Then, $B$ is of class $\mathcal{C}^\infty$, $L_g L_f^{i} B(\b{z}) = 0$ for $i\in\{0,1,2\}$, and
	\begin{equation}\label{eq:LgLf3B}
		L_gL_f^3B(\b{z}) = -\frac{\partial h_\mathrm{d}}{\partial\gamma}\,(\b{r}_3^\mathsf{T}\b{r}_{3d})\,\frac{\partial c}{\partial\b{p}_\delta}\left[(\b{r}_3^\mathsf{T}\b{r}_{3d})\,\b{I}+\b{S}(\b{e_r})\right]\b{e}_3\,,
	\end{equation}
	which is strictly negative for all $\b{z}$ with $\gamma<\gamma_f$ and $\b{r}_3^\mathsf{T}\b{r}_{3d}\geq\cos\theta_\mathrm{gs}$. Consequently, wherever these two conditions hold, $B$ has relative degree $\rho=4$ with respect to $\nu$, as required in Assumption~\ref{ass:ecbf}, and the ECBF-QP in \eqref{eq:ecbf_qp} is feasible.
\end{lemma}
\begin{pf}
	See Appendix~\ref{app:lie_derivatives}.
\end{pf}
Lemma~\ref{lemma:ecbf_is} makes Remark~\ref{rem:feasibility} explicit for this application. With $\b{r}_3=\b{r}_{3d}$, the derivation in Appendix~\ref{app:lie_derivatives} gives, for a generic desired path,
\begin{equation}\label{eq:LgLf3B_aligned}
	L_gL_f^3B=\frac{\partial c}{\partial\b{p}_\delta}\,\frac{\partial\b{p}_d}{\partial\gamma}\,,
\end{equation}
so the QP is feasible wherever the path is transversal to the level sets of $c$. For the vertical path \eqref{eq:nominal_path}, $\frac{\partial c}{\partial\b{p}_\delta}\b{e}_3=-\sin(\theta_\mathrm{gs})$ gives $L_gL_f^3B=\sin(\theta_\mathrm{gs})\,\partial h_\mathrm{d}/\partial\gamma$, so feasibility can only be lost if the attitude error exceeds $\theta_\mathrm{gs}$, or at $\gamma_f$, where $\partial h_\mathrm{d}/\partial\gamma=0$ ensures zero desired velocity at the landing site.

To ensure that the vector $\b{k}_\alpha$ satisfies the requirements of Lemma~\ref{th:ecbf}, we make use of Algorithm~\ref{alg:kalpha}. At the initial time, the closed-loop poles of the ECBF dynamics \eqref{eq:eta_b_dyn} under $\mu=-\b{k}_\alpha^\mathsf{T}\bs\eta_b(\b{z})$, given by $\bs{\lambda} = \left[\,\lambda_1\:\:\lambda_2\:\:\lambda_3\:\:\lambda_4\,\right]^\mathsf{T}$, are selected to satisfy the conditions stated in Lemma~\ref{th:ecbf}, with the functions $v_i$ in \eqref{eq:v_family} computed for $B$ in \eqref{eq:glide_ecbf} at $\b{z}(0)$. Then, $\b{k}_\alpha$ is computed by pole placement, so that $\b{A}_4-\b{B}_4\b{k}_\alpha^\mathsf{T}$ has eigenvalues $\bs\lambda$. Note that the pole selection depends on the adjustable scaling factors $a_i\mathrel{>}1$ and on the floor levels $b_i>0$. For $i=4$, $\dot v_3$ depends on $\nu$ and is evaluated with $\nu=\nu_\mathrm{nom}(0)$, so that $v_4(\b{z}(0))>0$ means that the nominal progression satisfies the ECBF constraint at $t=0$; by Lemma~\ref{th:ecbf}, any $\lambda_4<0$ would also be admissible.
\begin{algorithm}
\caption{Computation of $\b{k}_\alpha$}
\label{alg:kalpha}
\begin{algorithmic}[1]
\Require $B(\mathbf{z}(0))$, $a_i$, and $b_i$
\Ensure $\mathbf{k}_\alpha$ satisfies conditions of Lemma~\ref{th:ecbf}

\If{$t$ = 0}
    \State $v_0(\mathbf{z}(0)) \gets B(\mathbf{z}(0))$
    \For{$i = 1$ \textbf{to} $4$}
        \State $\lambda_i \gets -\max\left(-a_i\dfrac{\dot v_{i-1}(\mathbf{z}(0))}{v_{i-1}(\mathbf{z}(0))},\ b_i\right)$
        \State $v_i(\mathbf{z}(0)) \gets \dot v_{i-1}(\mathbf{z}(0)) - \lambda_i\, v_{i-1}(\mathbf{z}(0))$
    \EndFor
   \State $\mathbf{k}_\alpha \gets \mathrm{pole\_placement}(\bs{\lambda},\b{A}_4,\b{B}_4)$
\EndIf
\State \Return $\mathbf{k}_\alpha$
\end{algorithmic}
\end{algorithm}

By Lemmas~\ref{th:ecbf} and \ref{lemma:ecbf_is}, under the conditions of Lemma~\ref{lemma:ecbf_is}, the ECBF-QP is feasible and ensures $B(\b{z}(t))\geq0$, i.e., $\b{p}_\delta(t)\in\mathcal{S}$, as stated precisely in Theorem~\ref{thm:application}. The path governor is given by \eqref{eq:ecbf_qp_sol} with $\rho=4$ and $\b{k}_\alpha$ computed at the initial time by Algorithm~\ref{alg:kalpha}.
Since $\gamma$ nominally corresponds to the time variable, the nominal input \eqref{eq:nom_path_law} is obtained with $v_d\coloneq1$, i.e., $\overbar{\bs\gamma}_\mathrm{nom}=\left[\,1\:\:0\:\:0\,\right]^\mathsf{T}$, yielding
\begin{equation}\label{eq:nom_path_law_app}
\nu_\mathrm{nom} \coloneq  -k_{\gamma_1}\left(\dot{\gamma}-1\right) - k_{\gamma_2}\,\gamma^{(2)} - k_{\gamma_3}\,\gamma^{(3)} \,,
\end{equation}
where $\b{k}_\gamma = \left[\,k_{\gamma_1}\:\:k_{\gamma_2}\:\:k_{\gamma_3}\,\right]^\mathsf{T}$ is a constant gain vector such that $\b{A}_3-\b{B}_3\b{k}_\gamma^\mathsf{T}$ is Hurwitz. Note that $\nu^*$ enters the control law as a feedforward term that is not saturated, so the governor may demand arbitrarily large inputs to prevent constraint violation.

\begin{theorem}\label{thm:application}
	Consider the control-oriented model \eqref{eq:sys_design} in closed loop with the path following control law \eqref{eq:control_law}, the nominal path \eqref{eq:nominal_path}, and the ECBF-QP path governor \eqref{eq:ecbf_qp_sol}, with $\b{k}_\alpha$ computed by Algorithm~\ref{alg:kalpha}. Let $T_f$ be the first time at which $\gamma(T_f)=\gamma_f$, when the control is switched off ($T_f=\infty$ if this does not happen). Suppose that Assumption~\ref{ass:xi} holds, that $B(\b{z}(0))>0$, and that the attitude error remains below the glideslope angle, i.e.,
	\begin{equation}\label{eq:attitude_cond}
		\b{r}_3^\mathsf{T}(t)\,\b{r}_{3d}(t)\geq\cos\theta_\mathrm{gs}\,,
	\end{equation}
	for all $t\in[0,T_f)$. Then: (i) $\b{p}_\delta(t)\in\mathcal{S}$ for all $t\in[0,T_f]$, and $\nu$ departs from $\nu_\mathrm{nom}$ only when $\nu_\mathrm{nom}$ violates the ECBF condition; (ii) if $T_f<\infty$, then $\b{p}_\delta(T_f)-\b{p}_L=\b{e}_{\b{p}}(T_f)$ and $\b{v}_\delta(T_f)=\b{e}_{\b{v}}(T_f)$.

	If $\gamma$ and its first three time derivatives are bounded and the governor switches between active and inactive finitely many times, then: (iii) either $T_f<\infty$ and (ii) applies, or $\gamma(t)\to\gamma_f$, $\b{p}_\delta(t)\to\b{p}_L$, $\b{v}_\delta(t)\to\b{0}$ and, for almost every initial condition, $\b{r}_3(t)\to\b{e}_3$ as $t\to\infty$.
\end{theorem}
\begin{pf}
	(i) By Lemma~\ref{lemma:closed_loop}, the closed loop is in the form \eqref{eq:final_stab_sys}. For $t\in[0,T_f)$, $\gamma(t)<\gamma_f$ and \eqref{eq:attitude_cond} holds, so $L_gL_f^3B(\b{z}(t))<0$ by Lemma~\ref{lemma:ecbf_is}. Hence, the ECBF-QP is feasible, its solution is \eqref{eq:ecbf_qp_sol}, and $\mu\geq-\b{k}_\alpha^\mathsf{T}\bs\eta_b(\b{z})$ on $[0,T_f)$. Since $B(\b{z}(0))>0$, Algorithm~\ref{alg:kalpha} returns a $\b{k}_\alpha$ that satisfies the conditions of Lemma~\ref{th:ecbf}, so that $B(\b{z}(t))\geq0$, i.e., $\b{p}_\delta(t)\in\mathcal{S}$, for all $t\in[0,T_f)$, and at $T_f$ by continuity. Moreover, by \eqref{eq:ecbf_qp_sol}, $\nu=\nu_\mathrm{nom}$ whenever $\nu_\mathrm{nom}$ satisfies the ECBF condition.

	(ii) Since $\b{p}_d(\gamma_f)=\b{p}_L$ and $\partial\b{p}_d/\partial\gamma\,(\gamma_f)=\b{0}$, we have $\b{p}_d(\gamma(T_f))=\b{p}_L$ and $\dot{\b{p}}_d(T_f)=\b{0}$.

	(iii) As the governor switches finitely many times, there is $T\geq0$ such that it is either inactive on $[T,T_f)$ or active on $[T,T_f)$. Suppose that $T_f=\infty$. In the first case, $\nu=\nu_\mathrm{nom}$ on $[T,\infty)$, so $\dot\gamma\to1$ exponentially by Lemma~\ref{lemma:nominal_convergence}, and $\gamma$ reaches $\gamma_f$ in finite time, which contradicts $T_f=\infty$. In the second case, $\mu=-\b{k}_\alpha^\mathsf{T}\bs\eta_b(\b{z})$ on $[T,\infty)$, i.e., $\dot{\bs\eta}_b=(\b{A}_4-\b{B}_4\b{k}_\alpha^\mathsf{T})\bs\eta_b$, whose matrix is Hurwitz, so $B(\b{z}(t))\to0$. Since $\gamma,\dot\gamma,\ddot\gamma,\gamma^{(3)}$ are bounded{, $\nu$ is continuous,} and $h_\mathrm{d}$ is a polynomial, $t\mapsto\b{p}_d(\gamma(t))$ is of class $\mathcal{C}^4$ with bounded derivatives up to order three, and Proposition~\ref{prop:tracking} gives $\b{e}_{\b{p}}\to\b{0}$. As $B(\b{z})=c(\b{p}_\delta)$ and $\|\partial c/\partial\b{p}_\delta\|\leq{1}$ (see Appendix~\ref{app:lie_derivatives}), $|B(\b{z})-c(\b{p}_d(\gamma))|\leq{\|\b{e}_{\b{p}}\|}\to0$. Along the path \eqref{eq:nominal_path}, $c(\b{p}_d(\gamma))=\sin(\theta_\mathrm{gs})(h_\mathrm{d}(\gamma)-h_L)=\sin(\theta_\mathrm{gs})(h_0-h_L)(1-\gamma/\gamma_f)^2$, so that $(1-\gamma/\gamma_f)^2\to0$, i.e., $\gamma\to\gamma_f$. Then, $\b{p}_\delta=\b{p}_d(\gamma)+\b{e}_{\b{p}}\to\b{p}_L$ and $\b{v}_\delta=-\frac{\partial h_\mathrm{d}}{\partial\gamma}\dot\gamma\,\b{e}_3+\b{e}_{\b{v}}\to\b{0}$, since $\frac{\partial h_\mathrm{d}}{\partial\gamma}=\dot h_0(1-\gamma/\gamma_f)\to0$, $\dot\gamma$ is bounded, and $\b{e}_{\b{v}}\to\b{0}$. Finally, $\bs\xi-\bs\zeta=g\b{e}_3-\ddot{\b{p}}_d=\big(g+\frac{\partial^2h_\mathrm{d}}{\partial\gamma^2}\dot\gamma^2+\frac{\partial h_\mathrm{d}}{\partial\gamma}\ddot\gamma\big)\b{e}_3$, where $\frac{\partial^2h_\mathrm{d}}{\partial\gamma^2}>0$ and $\frac{\partial h_\mathrm{d}}{\partial\gamma}\ddot\gamma\to0$, while $\bs\zeta\to\b{0}$ by Proposition~\ref{prop:tracking}. Hence, $\b{r}_{3d}\to\b{e}_3$ and, consequently, $\b{r}_3\to\b{e}_3$ for almost every initial condition.
\end{pf}
\begin{remark}[Attitude condition]\label{rem:attitude}
	Condition \eqref{eq:attitude_cond} can be guaranteed from the initial condition. The time derivative \eqref{eq:Vdot_full} of the function $V$ in \eqref{eq:V_full} only requires Assumption~\ref{ass:xi} and holds for any progression $\gamma(t)$ of class $\mathcal{C}^4$, which includes the one produced by the governor, since \eqref{eq:ecbf_qp_sol} is continuous. Hence, $V$ is nonincreasing and $h_r(1-\b{r}_{3d}^\mathsf{T}\b{r}_3)\leq V(t)\leq V(0)$, i.e., $\b{r}_3^\mathsf{T}\b{r}_{3d}\geq1-V(0)/h_r$ for all $t\in[0,T_f)$. Therefore, \eqref{eq:attitude_cond} holds if the glideslope angle satisfies
	\begin{equation}\label{eq:theta_gs_cond}
		\cos\theta_\mathrm{gs}\leq1-\frac{V(0)}{h_r}\,,
	\end{equation}
	which can be checked at the initial time.
\end{remark}

\section{Experimental Validation}\label{sec:experiments}

The proposed approach is validated in an indoor flight arena with the \textit{QuadRocket} \cite{Santos2026}, a low-cost aerial platform that emulates a thrust-vectored rocket: a cylindrical rocket-like body is attached to a quadrotor through a universal joint, so that the body rotates freely about two axes while the quadrotor provides the thrust vector.

\begin{figure*}[t!]
	\centering
	\includegraphics{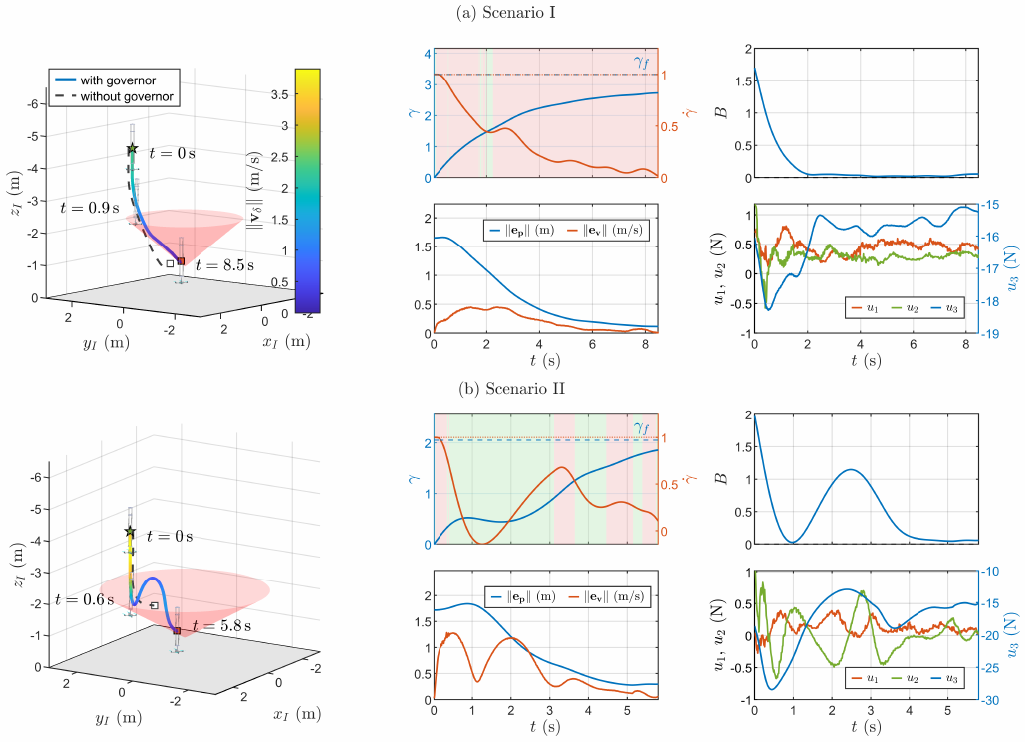}
\caption{Experimental landing results: (a) scenario I; (b) scenario II.
	On the left, trajectory of the control point $\b{p}_\delta$ with the governor, colored by its speed $\|\b{v}_\delta\|$ (common scale for both scenarios), with snapshots of the vehicle at the indicated instants, trajectory without the governor (dashed) up to the instant at which the metrics in Table~\ref{tab:landing} are evaluated, and boundary of the safe set $\mathcal{S}$ (red). On the right, evolution with the governor of (top left) the path parameter $\gamma$ and its rate $\dot\gamma$, with $\gamma_f$ (dashed) and the nominal rate $\dot\gamma=1$ (dotted), on a red background when the governor is active and a green background otherwise; (top right) the ECBF $B$; (bottom left) the norms of the position and velocity errors $\b{e}_{\b{p}}$ and $\b{e}_{\b{v}}$; and (bottom right) the components of the
	thrust-vector input $\b{u}$.}
	\label{fig:experimental_results}
\end{figure*}

\subsection{Experimental setup}

The experimental setup is the same as in \cite{Santos2026}. A Vicon motion capture system measures the poses of the body and of the quadrotor at 100~Hz, from which the state $(\b{p}_\delta,\b{v}_\delta,\b{r}_3,\bs{\Omega}_{12})$ is reconstructed in MATLAB/Simulink, with angular velocities obtained by first-order differentiation of the measured rotation matrices. The path following law \eqref{eq:control_law} and the path governor \eqref{eq:ecbf_qp_sol} run at 100~Hz, and the resulting thrust-vector command is converted into thrust and angular velocity commands, sent through an RC link to the onboard flight controller (Betaflight) of the quadrotor.

The vehicle parameters are $m=1.64$~kg, $j_\perp=0.15$~kg\,m$^2$, $\ell=0.5$~m and $\delta=0.18$~m. The path following law \eqref{eq:control_law} uses $k_1=2$, $\beta=0.3$, $k_I=1.8$, $k_2=1.5$, $s_1=1$, $s_2=3$, $s_3=1$, $k_r=60$, $h_r=6$, $k_\Omega=25$ and $h_\Omega=15$. The nominal path progression \eqref{eq:nom_path_law_app} uses $\b{k}_\gamma=[\,6\:\:11\:\:6\,]^\mathsf{T}$, and $\b{k}_\alpha$ is computed by Algorithm~\ref{alg:kalpha} with $a_i=2$ and $[\,b_1\:\:b_2\:\:b_3\:\:b_4\,]=[\,1\:\:1.5\:\:2\:\:2.5\,]$. The smoothing parameter of the glideslope constraint is $\epsilon=0.01$~m, and the apex of the glideslope cone is placed at the landing site $\b{p}_L$. These values are the same throughout the experiments.

Two landing scenarios were tested. In both, the landing controller is engaged when the control point crosses a trigger altitude $h_0$, and the nominal path \eqref{eq:nominal_path} is generated from the measured altitude and vertical velocity $\dot{h}_0$, which fixes $\gamma_f$ and the constant nominal deceleration $\ddot{h}_\mathrm{d}=\dot{h}_0^2/(2(h_0-h_L))$. The control is switched off manually by the operator, based on visual inspection, once the vehicle is close to the ground and essentially at rest. In scenario~I, the vehicle reaches $h_0=4.5$~m in a steady descent, with $\dot{h}_0=-2.16$~m/s, and the landing site is at $h_L=0.9$~m, with $\theta_\mathrm{gs}=50^\circ$, so that $\ddot{h}_\mathrm{d}=0.65$~m/s$^2$ and $\gamma_f=3.33$~s. In scenario~II, which is more aggressive, the vehicle starts from hover at 4.84~m, the controller is disabled and a constant thrust below the vehicle weight is commanded, emulating the final stage of an unpowered descent. The vehicle reaches $h_0=4.0$~m with $\dot{h}_0=-3.21$~m/s and, with $h_L=0.7$~m and $\theta_\mathrm{gs}=60^\circ$, the nominal path requires $\ddot{h}_\mathrm{d}=1.56$~m/s$^2$ and $\gamma_f=2.06$~s.

\subsection{Results and discussion}

The results are shown in Fig.~\ref{fig:experimental_results} and summarized in Table~\ref{tab:landing}, and a video of the flight experiments is available at \url{https://youtu.be/Rsbiqw4Nwi8}. Both scenarios were also flown without the governor, i.e., with $\nu=\nu_\mathrm{nom}$, from similar initial conditions. The metrics are evaluated at switch-off with the governor and, without it, when the control point crosses $h_L$ or $\gamma$ reaches $\gamma_f$, whichever occurs first; $e_h$ and $h-h_L$, with $h\coloneq-\b{e}_3^\mathsf{T}\b{p}_\delta$, are the horizontal distance and the height of the control point with respect to the landing site, and $t_\mathrm{land}$ is measured from engagement. With the governor, $B$ remains nonnegative in both scenarios, so the control point never leaves $\mathcal{S}$, and the vehicle approaches the landing site with near-zero velocity.

\begin{table}[!t]
\caption{Landing performance with and without the path governor}
\label{tab:landing}
\centering
\setlength{\tabcolsep}{4.5pt}
\begin{tabular}{llccccc}
\hline
 & Gov. & $\min B$ & $e_h$ & $h-h_L$ & $\|\b{v}_\delta\|$ & $t_\mathrm{land}$\\
 & & (m) & (m) & (m) & (m/s) & (s)\\
\hline
Sc.~I & with   & 0.02 & 0.11 & 0.16 & 0.01 & 8.48\\
      & without & $-0.33$ & 0.78 & 0.00 & 0.62 & 2.71\\
Sc.~II & with   & 0.03 & 0.24 & 0.20 & 0.03 & 5.80\\
      & without & 0.03 & 0.96 & 0.94 & 0.36 & 2.06\\
\hline
\end{tabular}
\end{table}

In scenario~I, the controller is engaged with a horizontal offset of about 1.6~m from the nominal path, which cannot be removed within the nominal descent. Without the governor, the control point leaves $\mathcal{S}$ before the end of the nominal path, and the vehicle hits the ground away from the landing site. With the governor, which is active almost throughout, $B$ decreases to nearly zero within 2~s, and the control point then slides along the boundary of the cone while the horizontal error decreases. Safety is obtained at the expense of descent speed, as $\dot\gamma$ decreases smoothly to nearly zero.

Since the governor acts only on $\gamma$, the desired path is the same with and without it, and only the instant at which each of its points is reached changes, as scenario~II illustrates. Without the governor, the constraint is satisfied along the nominal path, but, when $\gamma$ reaches $\gamma_f$, the control point is still about 1~m away from the landing site, both horizontally and vertically, and switching off the control would lead to a fall. With the governor, the progression is modified{, initially,} during the first 0.34~s, while $B$ is still above 1~m, about 0.65~s before $B$ reaches its minimum of 0.03~m. This anticipation follows from the ECBF condition, which accounts for the first three time derivatives of $B$, i.e., for how fast the boundary is approached. As $\nu$ is the fourth time derivative of $\gamma$, the intervention propagates through the chain of integrators. With the governor already inactive, $\dot\gamma$ keeps decreasing and briefly becomes negative, so that the desired position moves back up the path and the vehicle climbs away from the boundary, as shown by the growth of $B$, while $\dot\gamma$ recovers towards its nominal value, as predicted by Lemma~\ref{lemma:nominal_convergence}. Close to the landing site, where the cone narrows, the governor becomes intermittently active again.

At switch-off, with $\gamma<\gamma_f$ and $\dot\gamma$ close to zero, a horizontal path following error remains, due to effects not captured by \eqref{eq:sys_design}: the \textit{QuadRocket} is not a rocket-like vehicle in the sense of Definition~\ref{ass:rocket_like}, but a two-body system whose thrust vector is produced by a quadrotor, with actuator dynamics and delays. The integral action compensates these effects only slowly, since, once the velocity error vanishes, the integral error grows at the rate $\beta\bs\sigma(\b{e}_c,s_1)$, which is small for small errors. Since the apex of the cone is at $\b{p}_L$, a horizontal error $e_h$ is admissible only above the altitude $h_L+\cot(\theta_\mathrm{gs})\big(\sqrt{e_h^2+\epsilon^2}-\epsilon\big)$, i.e., 0.08~m and 0.13~m above the landing site, the remainder of $h-h_L$, scaled by $\sin(\theta_\mathrm{gs})$, being $B\approx{0.06}$~m. The governor therefore holds $\gamma$ back, consistently with Theorem~\ref{thm:application}, in which $\gamma$ approaches $\gamma_f$ only as the path following error vanishes, and placing $\b{p}_L$ in the interior of $\mathcal{S}$ would accommodate such errors. The constraint was nonetheless satisfied in both scenarios, indicating robustness to model uncertainty.

\section{High-fidelity Simulation}\label{sec:simulation}

To further assess the robustness of the method, a Monte Carlo study was conducted with the MATLAB/Simulink simulator of \cite{Santos2024sim}, extended for the landing phase with a throttleable engine and thrust misalignment, using the ode4 solver with a fixed step of 0.01~s. The model includes a rotating Earth with a zonal harmonic gravity model up to degree four (terms $J_2$ to $J_4$), the 1976 U.S. Standard Atmosphere, HWM14 wind with Dryden turbulence, tabulated aerodynamics, propellant-dependent mass properties, a first-order thrust lag of 0.1~s with limits of $[5,60]$~kN, and the gimbal dynamics of \cite{Santos2024sim} (0.02~s lag) with the deflection limit relaxed to $\pm45^\circ$, generous since only the glideslope constraint is considered. The spin is not regulated and is mostly determined by the initial condition. The vehicle is a finless LOX/RP-1 suborbital launcher, 8~m long, with a dry mass of 600~kg, 250~kg of propellant at the start of the landing burn, and the gimbal point at its base. The control law uses the instantaneous $m$, $j_\perp$ and $\ell$ (Assumption~\ref{ass:time_var}), while spin, aerodynamics, a moving center of mass, and thrust misalignment violate Assumptions~\ref{ass:spin} and \ref{ass:aero_free} and Definition~\ref{ass:rocket_like}.

The control law \eqref{eq:control_law} and the governor \eqref{eq:ecbf_qp_sol} were implemented with $k_1=0.4$, $\beta=3$, $k_I=0.03$, $k_2=0.5$, $s_1=2.5$, $s_2=5$, $s_3=4.5$, $k_r=100$, $h_r=30$, $k_\Omega=25$, $h_\Omega=15$, and $\b{k}_\gamma$ in \eqref{eq:nom_path_law_app}, $a_i$, $b_i$ as in Section~\ref{sec:experiments}, and $\theta_\mathrm{gs}=45^\circ$, $\epsilon=1$~m. The path \eqref{eq:nominal_path} starts at the initial altitude of the control point, and $\b{p}_L$ is placed 4.5~m above the landing pad to account for the distance between the control point and the base of the vehicle. A run stops when the base is 0.3~m above the ground or when $\gamma$ reaches $\gamma_f$, whichever occurs first. Each of 500 dispersed cases is simulated with and without the governor, with independent dispersions, where $\mathcal{U}[a,b]$ and $\mathcal{N}(0,\sigma^2)$ denote uniform and Gaussian distributions and directions are $\mathcal{U}[0,2\pi]$. The landing phase starts at $h_0=500\,\mathcal{U}[0.9,1.1]$~m with $\dot h_0=-100\,\mathcal{U}[0.9,1.1]$~m/s ($\gamma_f\in[8.3,12]$~s), a horizontal offset of magnitude $\mathcal{U}[0,150]$~m, horizontal velocity components $\mathcal{N}(0,5^2)$~m/s, a tilt of $\mathcal{U}[0,5]^\circ$, a transverse angular rate of $\mathcal{U}[0,2]^\circ$/s and a spin rate of $\mathcal{U}[-10,10]^\circ$/s. The dry mass and specific impulse are scaled by $\mathcal{U}[0.95,1.05]$, the propellant load and dry inertias by $\mathcal{U}[0.9,1.1]$, and the drag and lift coefficients by $\mathcal{U}[0.8,1.2]$; the dry center of mass and the center of pressure are shifted by $\mathcal{U}[-0.1,0.1]$ and $\mathcal{U}[-0.3,0.3]$~m; HWM14 is scaled by $\mathcal{U}[0,2]$, with a mean wind of $\mathcal{U}[0,10]$~m/s and moderate Dryden turbulence; and the thrust is misaligned by $\mathcal{U}[0,0.25]^\circ$.

\begin{figure}[!t]
	\centering
	\includegraphics{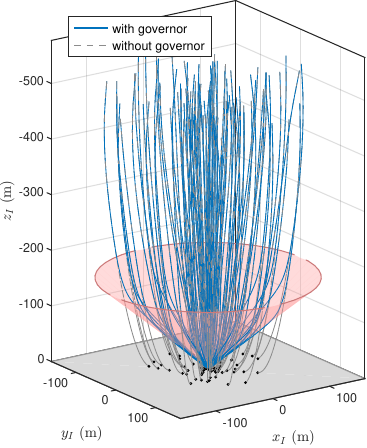}
	\caption{Simulated trajectories in 100 of the 500 paired runs, switch-off points (dots), and glideslope cone (red).}
	\label{fig:mc}
\end{figure}

\begin{table}[!t]
\caption{Monte Carlo results, 500 paired runs}
\label{tab:mc}
\centering
\setlength{\tabcolsep}{4.5pt}
\begin{tabular}{lcc}
\hline
Governor & without & with\\
\hline
Runs with $\min B<0$ & 458 & 286\\
$\min B$, worst run (m) & $-68.9$ & $-1.05$\\
$e_h$, median / max (m) & 16.9 / 98.9 & 1.26 / 2.22\\
$\|\b{v}_\delta\|$, median / max (m/s) & 5.74 / 14.7 & 0.47 / 0.77\\
$t_\mathrm{land}$, median (s) & 9.6 & 19.0\\
Propellant used, median (kg) & 59 & 90\\
\hline
\end{tabular}
\end{table}

Table~\ref{tab:mc} summarizes the results, with $e_h$ and $\|\b{v}_\delta\|$ the horizontal distance of the control point to the landing site and its speed at the end of the run (in the air for runs stopped at $\gamma_f$). Without the governor, the offset cannot be removed within the nominal descent, with the control point leaving the cone by more than 1~m in 416 runs, including 82 of the 166 with an initial horizontal offset below 50~m. Moreover, in 54 runs $\gamma$ reaches $\gamma_f$ with the base still up to 2~m above the ground. With the governor, the vehicle lands in all runs (Fig.~\ref{fig:mc}), with the control point within 2.22~m of the landing site and at less than 0.77~m/s, and $\gamma$ never reaches $\gamma_f$. $B$ becomes slightly negative in 286 runs, by more than 0.1~m in 174 and by at most about 1~m. Safety is again obtained at the expense of landing time and propellant, well within the available load. The hypotheses of Theorem~\ref{thm:application} hold in all runs, i.e., $B(\b{z}(0))\geq214$~m, $\|\bs{\xi}\|\geq9.8$~m/s$^2$, $\b{r}_3^\mathsf{T}\b{r}_{3d}\geq0.79>\cos\theta_\mathrm{gs}$, and the governor switches at most five times. The small violations are thus due to model mismatch, i.e., effects not captured by \eqref{eq:sys_design}. The thrust never saturates, and the gimbal angles saturate in 42 runs, with and without the governor, only during the initial attitude transient. The approach is thus robust to spin, aerodynamics and time-varying mass properties (Remark~\ref{rem:assumptions}).

\section{Conclusions}

This paper addressed the safe autonomous landing of rocket-like vehicles under a glideslope constraint by formulating the landing as a path following problem and using the progression of the path parameter to enforce the constraint. A framework was proposed for pre-stabilized systems in which an ECBF-based QP with a closed-form solution governs only the highest time derivative of the path parameter, departing from the nominal progression only when the latter would violate the ECBF condition and preserving the stability guarantees of the pre-stabilizing law. For a thrust-vector controlled rocket-like vehicle, a saturated nonlinear path following law and an ECBF of relative degree four for a smooth glideslope constraint were designed, and conditions were established under which the vehicle never leaves the safe set and reaches the landing site, in finite time up to the path following error or asymptotically. Flight experiments on the \textit{QuadRocket}, including an aggressive entry into the landing phase, showed that the glideslope constraint is satisfied in practice despite a significant mismatch between the platform and the design model{, and Monte Carlo simulations of a full-scale vehicle showed soft landings within the cone, up to sub-metre excursions, whereas most runs without the governor leave it}. Future work will address the main limitations of this first approach. Bounds on the thrust magnitude and pointing direction should be enforced together with the glideslope constraint, which requires guaranteeing the feasibility of a QP with several ECBF constraints. The nominal path could also be obtained from optimal guidance, for instance as a smoothed version of the last optimal trajectory, which requires a path of class $\mathcal{C}^4$ that lies strictly inside the safe set.

\begin{ack}
\sloppy
This work was supported by Fundação para a Ciência e a Tecnologia (FCT), through the Ph.D. scholarship of P. Santos (2023.00268.BD), LAETA (UID/50022/2025), and LARSyS (LA/P/0083/2020); by the Macao Science and Technology Development Fund under Grants 0192/2023/RIA3, 0059/2024/RIA1, 0036/2025/AIJ and 0021/2025/ASJ; and by the University of Macau, under Projects MYRG2022-00205-FST, MYRG-GRG2023-00107-FST-UMDF, UMDF-TISF/2025/007/FST, and SRG2024-00012-FST.
\end{ack}

\bibliographystyle{ieeetr}
\bibliography{autosam}

\appendix

\section{Proof of Proposition~\ref{prop:tracking}}\label{app:prop1}

By the assumptions of Proposition~\ref{prop:tracking}, $\b{r}_{3d}$, $\bs\Omega_{12c}$ and the control law are smooth in the closed-loop state $(\b{x},\b{e}_I)$, and $\b{r}_{3d}$ and $\bs\Omega_{12c}$ have derivatives bounded uniformly in $t$ on bounded sets{; $\b{p}_d^{(4)}$ enters the control law only through $\dot{\bs\Omega}_{12c}$, which cancels in the error dynamics below}. Using $\overline{u_3}=-\bs\xi^\mathsf{T}\b{r}_3$, $\b{S}^2(\b{r}_3)=\b{r}_3\b{r}_3^\mathsf{T}-\b{I}$ and \eqref{eq:ang_accel_law}, the closed loop reads
\begin{subequations}\label{eq:cl_err}
	\begin{align}
		\dot{\b{e}}_{\b{p}} &= \b{e}_{\b{v}}\,,\qquad
		\dot{\b{e}}_I = \beta\bs\sigma(\b{e}_c,s_1)+\b{e}_{\b{v}}\,,\\
		\dot{\b{e}}_{\b{v}} &= -\bs\zeta-\|\bs\xi\|\,\b{S}^2(\b{r}_3)\,\b{r}_{3d}\,,\label{eq:cl_err_v}\\
		\dot{\b{r}}_3 &= -\b{S}(\b{r}_3)\left(\b{e}_{\bs\Omega}+\bs\Omega_{12c}\right),\\
		\dot{\b{e}}_{\bs\Omega} &= -\tfrac{k_\Omega}{h_\Omega}\b{e}_{\bs\Omega}-\tfrac{h_r}{h_\Omega}\b{e_r}-\b{r}_3\,\dot{\b{r}}_3^\mathsf{T}\b{e}_{\bs\Omega}\,,\label{eq:cl_err_Om}
	\end{align}
\end{subequations}
where the last term of \eqref{eq:cl_err_Om} is $-(\b{I}+\b{S}^2(\b{r}_3))\dot{\bs\Omega}_{12c}$, rewritten by differentiating $\b{r}_3^\mathsf{T}\bs\Omega_{12c}=0$. Consider now the Lyapunov function
\begin{equation}\label{eq:V_full}
	V \coloneq V_2 + h_r\left(1-\b{r}_{3d}^\mathsf{T}\b{r}_3\right) + \frac{h_\Omega}{2}\|\b{e}_{\bs{\Omega}}\|^2\,.
\end{equation}
Since $\frac{\mathrm{d}}{\mathrm{d}t}h_r(1-\b{r}_{3d}^\mathsf{T}\b{r}_3)=h_r\b{e_r}^\mathsf{T}(\bs\Omega_{12}-\bs\Omega_{12d})$, substituting $\bs\Omega_{12}=\b{e}_{\bs\Omega}+\bs\Omega_{12c}$ and using $\b{S}^2(\b{r}_3)\b{e_r}=-\b{e_r}$, $\b{e_r}^\mathsf{T}\b{S}(\b{r}_3)\dot{\b{e}}_I=\dot{\b{e}}_I^\mathsf{T}\b{S}^2(\b{r}_3)\b{r}_{3d}$ (which cancels $\Delta$ in \eqref{eq:v2_dot_2}) and $\b{r}_3^\mathsf{T}\b{e}_{\bs\Omega}=0$, we obtain
\begin{equation}\label{eq:Vdot_full}
	\dot{V} = -W_1 - k_r\|\b{e_r}\|^2- k_\Omega\|\b{e}_{\bs{\Omega}}\|^2 \leq -\b{e}^\mathsf{T}\b{Q}\b{e}\leq0\,,
\end{equation}
with $\b{e} \coloneq \big[\bs{\sigma}^\mathsf{T}(\b{e}_c,s_1)\;\bs{\sigma}^\mathsf{T}(\dot{\b{e}}_I,s_3)\;\b{e_r}^\mathsf{T}\;\b{e}_{\bs{\Omega}}^\mathsf{T}\big]^\mathsf{T}$ and $\b{Q}\coloneq \beta\b{I} \oplus k_2\b{I} \oplus k_r\b{I} \oplus k_\Omega\b{I}$, where $\b{x}^\mathsf{T}\bs\sigma(\b{x},s)\geq\|\bs\sigma(\b{x},s)\|^2$ was used. As $V$ grows unboundedly with $\b{e}_c$, $\b{e}_{\b{v}}$, $\b{e}_I$ and $\b{e}_{\bs\Omega}$, $V(t)\leq V(0)$ bounds all errors and $\bs\Omega_{12}$, so solutions exist on $[0,\infty)$. By \eqref{eq:Vdot_full}, $\int_{0}^\infty\b{e}^\mathsf{T}\b{Q}\b{e}\,\mathrm{d}t\leq V(0)$, and $\dot{\b{e}}$ is bounded, so Barbalat's lemma \cite[Lemma~8.2]{Khalil} gives $\b{e}\to\b{0}$. Thus $\b{e}_c,\dot{\b{e}}_I\to\b{0}$, hence $\b{e}_{\b{v}}=\dot{\b{e}}_I-\beta\bs\sigma(\b{e}_c,s_1)\to\b{0}$ and $\b{e}_{\b{p}}\to\b{0}$. As $\ddot{\b{e}}_{\b{v}}$ is bounded, Barbalat's lemma also gives $\dot{\b{e}}_{\b{v}}\to\b{0}$, and, since $\|\b{S}^2(\b{r}_3)\b{r}_{3d}\|=\|\b{e_r}\|\to0$, \eqref{eq:cl_err_v} yields $\bs\zeta\to\b{0}$, i.e., $\b{e}_I\to\b{0}$. Finally, $(\b{r}_{3d}^\mathsf{T}\b{r}_3)^2=1-\|\b{e_r}\|^2\to1$, so, by continuity, $\b{r}_{3d}^\mathsf{T}\b{r}_3$ converges to $+1$ or to $-1$.

\section{Proof of Lemma~\ref{lemma:ecbf_is}}   \label{app:lie_derivatives}

Since $\epsilon>0$, the function $c$ in \eqref{eq:smooth_slope}, and hence $B$, is of class $\mathcal{C}^\infty$, with
\begin{gather*}
	\frac{\partial c}{\partial\b{p}_\delta}=-\sin(\theta_\mathrm{gs})\,\b{e}_3^\mathsf{T}-\cos(\theta_\mathrm{gs})\,\b{q}^\mathsf{T},\\
	\b{q}\coloneq\frac{-\b{S}^2(\b{e}_3)\overbar{\b{p}}_\delta}{\|{-}\b{S}^2(\b{e}_3)\overbar{\b{p}}_\delta+\epsilon\,\b{e}_3\|}\,,
\end{gather*}
where $\b{e}_3^\mathsf{T}\b{q}=0$ and $\|\b{q}\|<1$, so that $\frac{\partial c}{\partial\b{p}_\delta}\b{e}_3=-\sin(\theta_\mathrm{gs})$ and $\|\partial c/\partial\b{p}_\delta\|\leq1$.

\textit{Relative degree.} As $B$ depends on $\b{z}$ only through $\b{p}_\delta$, the chain rule gives $B^{(k)}=\frac{\partial c}{\partial\b{p}_\delta}\b{p}_\delta^{(k)}$ plus terms in $\b{p}_\delta^{(1)},\ldots,\b{p}_\delta^{(k-1)}$. Hence, $\nu$ appears in $B$ for the first time at the order at which it appears in $\b{p}_\delta$, and then $L_gL_f^{k-1}B=\frac{\partial c}{\partial\b{p}_\delta}\,\partial\b{p}_\delta^{(k)}/\partial\nu$. By Lemma~\ref{lemma:closed_loop}, $\nu$ drives only $\bs\Omega_{12}$ and $\gamma^{(3)}$. Neither $\dot{\b{p}}_\delta=\b{v}_\delta$ nor $\ddot{\b{p}}_\delta=g\b{e}_3-\b{r}_3\b{r}_3^\mathsf{T}\bs\xi$ depends on them, since $\bs\xi$ depends on $\b{p}_\delta$, $\b{v}_\delta$, $\gamma$, $\dot\gamma$, $\ddot\gamma$, and on $\b{e}_I$, whose dynamics do not depend on $\nu$. Thus, $\b{p}_\delta^{(3)}$ is free of $\nu$ and $L_gL_f^iB=0$ for $i\in\{0,1,2\}$. In turn,
\begin{equation*}
	\b{p}_\delta^{(3)}=-(\dot{\b{r}}_3\b{r}_3^\mathsf{T}+\b{r}_3\dot{\b{r}}_3^\mathsf{T})\bs\xi-\b{r}_3\b{r}_3^\mathsf{T}\dot{\bs\xi}
\end{equation*}
depends on $\bs\Omega_{12}$ through $\dot{\b{r}}_3=-\b{S}(\b{r}_3)\bs\Omega_{12}$, and on $\gamma^{(3)}$ through $\dot{\bs\xi}\ni-\dddot{\b{p}}_d\ni\frac{\partial h_\mathrm{d}}{\partial\gamma}\gamma^{(3)}\b{e}_3$. With $\b{g}_{\bs\Omega}$ in \eqref{eq:g_Omega},
\begin{equation*}
	\frac{\partial\b{p}_\delta^{(4)}}{\partial\nu}=\left[(\b{r}_3^\mathsf{T}\bs\xi)\b{S}(\b{r}_3)+\b{r}_3\bs\xi^\mathsf{T}\b{S}(\b{r}_3)\right]\b{g}_{\bs\Omega}-\frac{\partial h_\mathrm{d}}{\partial\gamma}(\b{r}_3^\mathsf{T}\b{e}_3)\,\b{r}_3\,.
\end{equation*}
Using $\b{S}^3(\b{r}_3)=-\b{S}(\b{r}_3)$, $\bs\xi=\|\bs\xi\|\b{r}_{3d}$, $\b{S}(\b{a})\b{S}(\b{b})=\b{b}\b{a}^\mathsf{T}-(\b{a}^\mathsf{T}\b{b})\b{I}$, and $\b{r}_{3d}\b{r}_3^\mathsf{T}-\b{r}_3\b{r}_{3d}^\mathsf{T}=-\b{S}(\b{e_r})$, this simplifies to
\begin{equation*}
	\frac{\partial\b{p}_\delta^{(4)}}{\partial\nu}=-\frac{\partial h_\mathrm{d}}{\partial\gamma}\,(\b{r}_3^\mathsf{T}\b{r}_{3d})\left[(\b{r}_3^\mathsf{T}\b{r}_{3d})\,\b{I}+\b{S}(\b{e_r})\right]\b{e}_3\,,
\end{equation*}
which, multiplied by $\frac{\partial c}{\partial\b{p}_\delta}$, yields \eqref{eq:LgLf3B}. The derivation only uses that $\nu$ enters through $\b{p}_d^{(4)}$, with coefficient $\partial\b{p}_d/\partial\gamma=-\frac{\partial h_\mathrm{d}}{\partial\gamma}\b{e}_3$; for a generic path, $-\frac{\partial h_\mathrm{d}}{\partial\gamma}\b{e}_3$ is replaced by $\partial\b{p}_d/\partial\gamma$, which gives \eqref{eq:LgLf3B_aligned} for $\b{r}_3=\b{r}_{3d}$.

\textit{Sign.} Let $\varphi\in[0,\pi]$ be the angle between $\b{r}_3$ and $\b{r}_{3d}$, so that $\b{r}_3^\mathsf{T}\b{r}_{3d}=\cos\varphi$ and $\|\b{e_r}\|=\sin\varphi$. As $\b{e}_3^\mathsf{T}\b{S}(\b{e_r})\b{e}_3=0$,
\begin{equation*}
	L_gL_f^3B=\frac{\partial h_\mathrm{d}}{\partial\gamma}\cos\varphi\left[\sin\theta_\mathrm{gs}\cos\varphi+\cos\theta_\mathrm{gs}\,\b{q}^\mathsf{T}\b{S}(\b{e_r})\b{e}_3\right],
\end{equation*}
with $|\b{q}^\mathsf{T}\b{S}(\b{e_r})\b{e}_3|\leq\|\b{q}\|\sin\varphi$. For $\varphi\leq\theta_\mathrm{gs}<\pi/2$, the bracket is at least $\sin(\theta_\mathrm{gs})\cos\varphi-\cos(\theta_\mathrm{gs})\|\b{q}\|\sin\varphi$, which equals $\sin\theta_\mathrm{gs}$ for $\varphi=0$ and exceeds $\sin(\theta_\mathrm{gs}-\varphi)\geq0$ otherwise. Moreover, $\frac{\partial h_\mathrm{d}}{\partial\gamma}=\dot h_0(1-\gamma/\gamma_f)<0$ for $\gamma<\gamma_f$. Hence, $L_gL_f^3B<0$ whenever $\gamma<\gamma_f$ and $\b{r}_3^\mathsf{T}\b{r}_{3d}\geq\cos\theta_\mathrm{gs}$, so the ECBF-QP is feasible there. Finally, $L_f^4B$ in \eqref{eq:ecbf_qp_sol} is $B^{(4)}$ evaluated with $\nu=0$, and is computed symbolically in the implementation.

\end{document}